\documentclass[aps,prd,reprint,superscriptaddress,nofootinbib,longbibliography]{revtex4-2}

\usepackage[T1]{fontenc}
\usepackage{lmodern} 
\usepackage[utf8]{inputenc}
\usepackage{amsmath,amssymb,amsthm,mathtools,bm}
\usepackage{graphicx}
\usepackage{booktabs,multirow}
\usepackage{enumitem}
\usepackage[protrusion=true,expansion=false]{microtype}
\usepackage{xcolor}
\usepackage{tikz}
\usepackage[colorlinks=true,linkcolor=blue!55!black,
            citecolor=green!40!black,urlcolor=blue!55!black]{hyperref}

\allowdisplaybreaks
\graphicspath{{figures/}}

\newcommand{\paperfigure}[3][\linewidth]{%
  \IfFileExists{#3}{\includegraphics[width=#1]{#3}}{%
    \IfFileExists{figures/#3}{\includegraphics[width=#1]{figures/#3}}{%
      \fbox{\parbox[c][#2][c]{\dimexpr#1-2\fboxsep-2\fboxrule\relax}{%
        \centering\normalfont\small Figure file not supplied\\[6pt]
        \texttt{\detokenize{#3}}\\[6pt]
        \footnotesize Layout placeholder}}}}}

\numberwithin{equation}{section}

\newcommand{\dd}{\mathrm{d}}
\newcommand{\Mp}{M_{\rm Pl}}
\newcommand{\rDE}{\rho_{\rm DE}}

\newcommand{\Ode}{\Omega_{\rm DE}}
\newcommand{\Om}{\Omega_m}
\newcommand{\LCDM}{\Lambda\mathrm{CDM}}
\newcommand{\rinf}{r_\infty}

\newcommand{\kms}{{\rm km\,s^{-1}\,Mpc^{-1}}}

\theoremstyle{definition}

\theoremstyle{remark}

\usepackage{comment}
\begin{document}

\title{{\LARGE Relaxation Dark Energy:}\\
\vskip0.2cm
{\Large Compositional Dynamics and the Future of Cosmic Acceleration}}
\author{Gianmassimo Tasinato}
\affiliation{Department of Physics, Swansea University, Swansea SA2 8PP, United Kingdom,} \affiliation{Dipartimento di Fisica e Astronomia, Universit\`a di Bologna, Via Irnerio 46, 40126 Bologna, Italy,\\
{\bf{ email}:} {g.tasinato2208.at.gmail.com}
\bigskip\medskip
}

\begin{abstract}
\noindent
If cosmic acceleration is driven by a positive cosmological constant, the universe approaches an eternally accelerating de Sitter state, whose interpretation in quantum gravity remains elusive. We develop a framework in which acceleration is instead a transient stage of dark-sector relaxation towards compositional equilibrium: matter and dark energy approach a stationary density ratio, both eventually diluting like pressureless matter. Guided by linear response theory and Onsager’s near-equilibrium regression hypothesis, we describe this evolution through a composition variable measuring departure from that state. Adopting constant fractional relaxation as a global background law yields a dark-energy equation of state \(w=w_0a^{-n}\), with \(n\) interpreted as a relaxation rate per e-fold of expansion. This history connects early dark-energy suppression, an upward phantom crossing and a density maximum to a pressureless scaling future. Using DESI DR2 BAO and Planck distance priors, with and without three alternative supernova compilations, we find a nominal preference for nonzero relaxation over constant $w$ at $2.6$--$3.6\sigma$. The composition description also motivates a simple constitutive prescription for linear perturbations, whose evolution equations remain regular at phantom crossing. Near equilibrium, regular long-wavelength composition disturbances relax at the background rate to leading order, while small-scale composition fluctuations propagate as regular damped sound waves. We prove, however, that with separately conserved matter and dark energy, our pointwise constitutive relation cannot support a regular, gradient-stable negative-pressure branch throughout its full past evolution. We explore modifications of the pressure response and their consistency with neighbouring homogeneous histories, identifying requirements for extending the near-equilibrium description to the full cosmological evolution.
\end{abstract}

\maketitle

\section{Introduction}
\label{sec:introduction}

\begingroup
\renewcommand{\thefootnote}{\fnsymbol{footnote}}
\begin{flushright}
  \textit{Vedi che gi\`a non se' n\'e due n\'e uno.}%
  ~\footnote[1]{Dante Alighieri, \textit{Divina Commedia}:
  \textit{Inferno}, XXV, 69.}
\end{flushright}
\endgroup

\medskip
\noindent
What is the future of our universe? If cosmic acceleration is driven
by a positive cosmological constant, the universe approaches an
eternally accelerating de Sitter state, while the ratio of dark
energy (DE) to matter grows without bound. Such a scenario
is problematic for quantum gravity, as first pointed out in \cite{Witten:2001kn,FischlerR,HellermanR} -- see e.g. 
\cite{AndriotR} for a recent review. Dynamical DE allows a
different possibility: the present accelerated epoch may be a
transient stage in an evolution towards stationary dark-sector
composition. Matter and DE would then approach a finite density
ratio and dilute at the same rate, bringing acceleration to an end
in the distant future. In this work we investigate this possibility
as a process of \emph{compositional relaxation}. Guided by linear
response theory and Onsager's near-equilibrium regression
hypothesis, we connect the relaxation of the background composition
to the response of its perturbations. We explore how this description
can be extended beyond the neighbourhood of equilibrium, and ask
whether its background relaxation rate can be constrained by
current observations.

\smallskip

This question is timely in light of indications for evolving DE
in DESI-era analyses~\cite{DESIDR2A}. The
Chevallier--Polarski--Linder (CPL) parametrization,
$w_{\rm CPL}(a)=w_0+w_a(1-a)$, provides a useful description of
the expansion history over the observed redshift range. Several
data combinations favour histories crossing the phantom divide,
$w=-1$, although the evidence depends on the datasets and modelling
assumptions. Such parametrizations describe the observed evolution, but
without identifying a future physical endpoint. The relaxation description
we propose instead connects the measured history to a state of
stationary composition and asks what physical response could drive
the approach to that state.

A concrete motivation comes from the DE equation of state
$w=w_0a^{-1/2}$, with $w_0<0$, identified through a data-driven
analysis within VCDM~\cite{Borghetto2026}. For separately conserved
DE and pressureless matter, the DE density and relative abundance
are
\begin{align}
  \rho_{\rm DE}(a)
  &=\rho_{{\rm DE},0}a^{-3}
    \exp\!\left[6w_0(a^{-1/2}-1)\right],
    \nonumber\\
  \frac{\rho_{\rm DE}(a)}{\rho_m(a)}
  &=r_0\exp\!\left[6w_0(a^{-1/2}-1)\right],
  \label{eq:rPowerA}
\end{align}
where $a=1$ today and
$r_0=\rho_{{\rm DE},0}/\rho_{m,0}$.
DE is exponentially suppressed at early times, grows during its
phantom phase, and reaches a maximum density when it crosses
$w=-1$ upwards. Its density subsequently decreases, while its
abundance relative to matter continues to increase towards a finite
limit. At late times $w\to0$, and both components dilute like
pressureless matter. For histories reproducing present acceleration,
the accelerated phase therefore ends. These features suggest that the two dilution rates become equal as the
composition approaches a stationary state. Similar behaviours were
found in \cite{Barrow:2000nc,Sen:2005sk}.

The explicit model in \cite{Borghetto2026}
can be seen as particular case of a broad family of dark
energy models exhibiting relaxation. 
To pursue this interpretation 
the choice of relaxation variable is central.
In an expanding universe, the decrease of an individual density
does not by itself measure approach to a stationary composition.
Assuming separate conservation of DE and pressureless matter, the
ratio $r=\rho_{\rm DE}/\rho_m$ removes their common $a^{-3}$
dilution and isolates the difference between their dilution rates.
On a negative-pressure branch, $r'/r=-3w>0$, so the relative
abundance increases monotonically. Its logarithm accommodates
changes spanning many orders of magnitude and turns fractional
changes in abundance into additive displacements. When a finite
asymptotic ratio $r_\infty$ exists --  and we assume it does -- the \emph{composition variable}
$X=\ln(r/r_\infty)$ measures departure from that ratio.
Separate conservation gives the exact identity $X'=-3w$, where a
prime denotes $d/d\ln a$, directly connecting composition evolution
to the observable equation of state.

On a DE negative-pressure branch, $X$ is monotonic and the history
can be represented by a first-order autonomous flow,
$X'=-\Gamma(X)$, with $\Gamma=3w$ playing the role of an effective
force. The composition--force plane places phantom crossings,
density extrema and scaling endpoints within a common geometric
description. A zero $\Gamma(0)=0$ with $\Gamma_X(0)>0$ describes
an attractive state of stationary relative abundance. We call this state
\emph{compositional equilibrium}. The individual densities continue
to dilute, and thermodynamic equilibrium is not assumed.

Assuming that compositional equilibrium exists, 
near it the system response
has the form $\Gamma(X)=nX+ {\cal O}(X^2)$ with $n>0$.
This provides a natural setting for relating the relaxation of a
small background displacement to the regression of composition
disturbances. Our simplest additional hypothesis extends the
linear law to the whole background trajectory -- so that the
logarithmic displacement relaxes at a constant fractional rate
per e-fold of expansion. This gives $\Gamma=nX$ and the DE
equation of state $w=w_0a^{-n}$. Its global validity is a 
testable  assumption.  The
motivating square-root history of Eq.~\eqref{eq:rPowerA} is the
benchmark $n=1/2$.

Relative abundances and scaling solutions have substantial
precedent. Dalal et al.~\cite{DalalR} parametrized relative scaling;
interacting-quintessence models developed finite-ratio
attractors~\cite{ZimdahlPavonChimentoR,ChimentoR,ZimdahlPavonScalingR};
and von Marttens et al.~\cite{vonMarttensR} organized interacting
models through functions of the density ratio. State-space
diagnostics likewise connect the equation of state to physical
evolution~\cite{CaldwellLinderR,SahniOmR}, and the form
$w=w_0a^{-n}$ has been studied phenomenologically by
Hossain~\cite{HossainR}. Other papers discussing
versions of composition dynamics include \cite{JungleR,JungleRefR,SoftCoincidenceR,ArevaloR,FunoR,VeltenR,KaloperOliveR,PanR,CaoR,AmendolaR,WangReviewR,BolotinR,DarkDegeneracyR,KunzR}.
Our contribution is to identify a
relaxation principle for selecting these histories, give their
parameters a physical interpretation, and -- above all -- use the same
compositional description to connect background evolution with
the dynamics of cosmological perturbations.

We confront linear relaxation with DESI DR2 BAO measurements,
Planck distance priors and three supernova compilations. Within
these background likelihoods, a nonzero rate is preferred over the
constant-$w$ limit at a nominal significance of
$2.6$--$3.6\sigma$. The general relaxation family fits comparably
to CPL, while fixing $n=1/2$ retains comparable fit quality with
one fewer DE parameter. We also test the defining constant-rate
property by allowing the logarithmic slope of $|w|$ to vary.
Present data show no significant departure from constancy,
although the test has limited sensitivity. The fitted histories
predict a quantifiable  recent DE density maximum and an eventual end of
acceleration -- the latter being a conditional extrapolation of the
assumed global law, rather than a directly measured future epoch.

The description of perturbation dynamics requires additional physical
input. Effective phantom crossing can arise in distinct settings,
including quintom models~\cite{Feng2005}, effective fluids and
parametrized post-Friedmann
descriptions~\cite{Hu2005,KunzSapone2006,FangHuLewis2008},
interacting sectors~\cite{AmendolaR,WangReviewR,Valiviita2008},
and modified or constrained gravity, including
VCDM~\cite{MukohyamaNoui2019,DeFeliceDollMukohyama2020}.
Other works related to the phantom-crossing subject include \cite{OzulkerR,SilvaNunesR,ChenClineR,YeR,DEPerturbationsRevisitedR,AmendolaTsujikawaR}.
Regular crossing of a background equation of state does not
establish a healthy microscopic realization, as illustrated by
single-field crossing obstructions, see e.g.~\cite{Vikman2005,Creminelli:2008wc}. For a genuine
positive-density material fluid, a phantom phase violates its
individual null energy condition; preservation of the total NEC
does not establish the stability of its fluctuations. Different
realisations of the same composition history can therefore predict
dramatically different growth and lensing signals.

However,
near compositional equilibrium, our framework supplies a concrete
and physically well-motivated 
connection between background relaxation and fluctuation response.
Promoting the composition law to a local constitutive relation
gives long-wavelength composition disturbances the same relaxation
rate as the background, realizing an Onsager-type  near-equilibrium
regression principle. Relative abundance disturbances also have a
positive acoustic response, while in the scaling limit common
matter--DE disturbances can cluster gravitationally. The resulting
description therefore combines relaxation of composition with the
persistence of structure. 

These properties motivate investigating how the constitutive
description can be extended to backgrounds farther from equilibrium.
The gauge-invariant composition perturbation remains regular at
phantom crossing, where the conventional relative entropy variable
degenerates. Regularity alone does not ensure stability, however:
we prove that imposing the background pressure law pointwise can not
maintain a gradient-stable negative-pressure branch throughout the
past. This restriction identifies where additional response physics
is required beyond the near-equilibrium description. We examine
pressure modifications, their separate-universe limitations, and
the conditions that a consistent physical completion must satisfy.

The paper is organized as follows. Section~\ref{sec:state-space} develops the
composition-flow description and its relation to phantom crossing,
acceleration and the asymptotic fate of the universe. Section~\ref{sec:relaxation-model}
formulates linear relaxation and derives its background
consequences. Section~\ref{sec:observables} presents the observational constraints
and the physical epochs implied by the fitted histories.
Section~\ref{sec_cosflu} studies composition perturbations, their near-equilibrium
response, the obstruction to a stable pointwise constitutive relation, and
possible extensions. Section~\ref{sec:discussion} outlines the theoretical and
observational directions opened by this framework.
Appendices~\ref{app:method} and \ref{app_null} explain the numerical implementation and the
direct null test of constant relaxation.

\section{Setup for composition space}
\label{sec:state-space}

The expansion history 
records how the universe evolved so far. But it does not by itself
suggest where the evolution is heading.  In this work we explore
the possibility that the universe is on
its way towards a compositional equilibrium and 
we theoretically motivate a simple parametrization
for the evolution of the dark sector.
We test our hypothesis with current data, and
we discuss its consequences for the evolution
of fluctuations. 

\smallskip

Compositional equilibrium is defined as
a final state in which (evolving) dark energy and pressureless matter -- including dark matter --
scale with the same rate.  It requires us to think and describe the two components of the dark sector as following a concerted evolution towards equilibrium. 
For this purpose,  in this Section
we  introduce and study the cosmological evolution in terms of   an appropriate {\it composition variable}, called $X$, which is a function of the ratio of dark energy versus pressureless matter densities.  We show 
that the  evolution
of the dark sectors -- when expressed in terms of $X$ --  can be fully described in terms of a first order  autonomous  equation. From the structure of such equation  we can read several physical
properties of the cosmological history, depending on the features of a single flow curve spanning a one-dimensional direction. In particular,
phantom crossing, extrema of the dark-energy density, and the  approach to final a compositional equilibrium 
 have all specific and simple 
geometrical meanings in composition space. Our aim
in this Section is to  familiarize
the reader with the physics of the compositional  sector and how it approaches equilibrium --  before
selecting a physically well-motivated relaxation  framework -- Section~\ref{sec:relaxation-model} --   test it with
 current observational
data -- Section~\ref{sec:observables} --  and elaborate on the dynamics
of its cosmological fluctuations  --  Section~\ref{sec_cosflu}.

\subsection{Composition variable and its autonomous first-order equation.}
\label{sub_compo}

We consider an expanding, spatially flat FLRW background containing pressureless
matter and a separately conserved effective dark component, with positive
energy densities. Their homogeneous evolution obeys
\begin{equation}
\begin{aligned}
 3\Mp^2H^2 &= \rho_m+\rDE,\\
 \rho_m'&=-3\rho_m,\\
 \rDE'&=-3(1+w)\rDE.
\end{aligned}
 \label{eq:background}
\end{equation}
where $w$ is the effective DE equation of state and a prime denotes
$\dd/\dd\ln a$. 
Here $\rho_m$ denotes the total pressureless component, cold dark matter plus baryons. 
These equations specify the background description in terms of a coarse-grained fluid  prescription. They do
not commit on  the microscopic origin or perturbative properties of the dark
component. We assume a separate conservation of dark
matter and dark energy at the background level -- as in Eq.~\eqref{eq:background}. This 
is an important property  for our arguments, and sets a difference with respect to other scenarios.

Within appropriate and  physically well motivated hypothesis, we can re-express the evolution equations for the dark sector system in terms of a single autonomous first order equation.
Changes in the relative
abundance arise from different dilution rates, without requiring direct energy transfer
between the two homogeneous sectors.
Introducing
\begin{equation}
 r\equiv\frac{\rDE}{\rho_m},
 \qquad
 \Ode=\frac{r}{1+r},
 \label{eq:rOmega}
\end{equation}
we obtain
\begin{equation}
\begin{aligned}
 \frac{r'}{r}&=-3w,\\
 \Ode'&=-3w\Ode(1-\Ode),\\
 \frac{H'}{H}&=-\frac32(1+w\Ode).
\end{aligned}
 \label{eq:master}
\end{equation}
The first of these relations isolates the physical origin of compositional evolution.
The common dilution due to expansion cancels in the ratio, leaving only the
effect of the dark component's pressure. When $w<0$, dark energy gains in
relative abundance. This remains true even when its own density decreases:
for $-1<w<0$, both densities fall, but matter falls faster. The last relation
in Eq.~\eqref{eq:master} describes a different aspect of the evolution. The
expansion responds to the pressure weighted by the abundance of the component
carrying it -- a distinction that will be essential when discussing acceleration.

But for our study 
it is more convenient to
introduce  a logarithmic coordinate, the composition variable $X$,  which    measures fractional changes in
composition. It accommodates histories spanning many orders of magnitude in
$r$, and obeys a simple evolution equation. We  define it relative to an arbitrary fixed positive reference ratio,
\begin{equation}
 X\equiv\ln\frac{r}{r_{\rm ref}},
 \label{eq:Xdef}
\end{equation}
so that separate conservation gives the exact identity
\begin{equation}
 X'=-3w.
 \label{eq:Xidentity}
\end{equation}
Changing $r_{\rm ref}$ only translates the origin of $X$; it does not change the
physical trajectory.  $X$ is the central
quantity in our arguments -- it is
the coarse-grained, slow composition variable that approaches
 equilibrium, within a  framework of relaxing dark sector we are going to develop. 

\begin{figure}[t]
\centering
\paperfigure{1.75in}{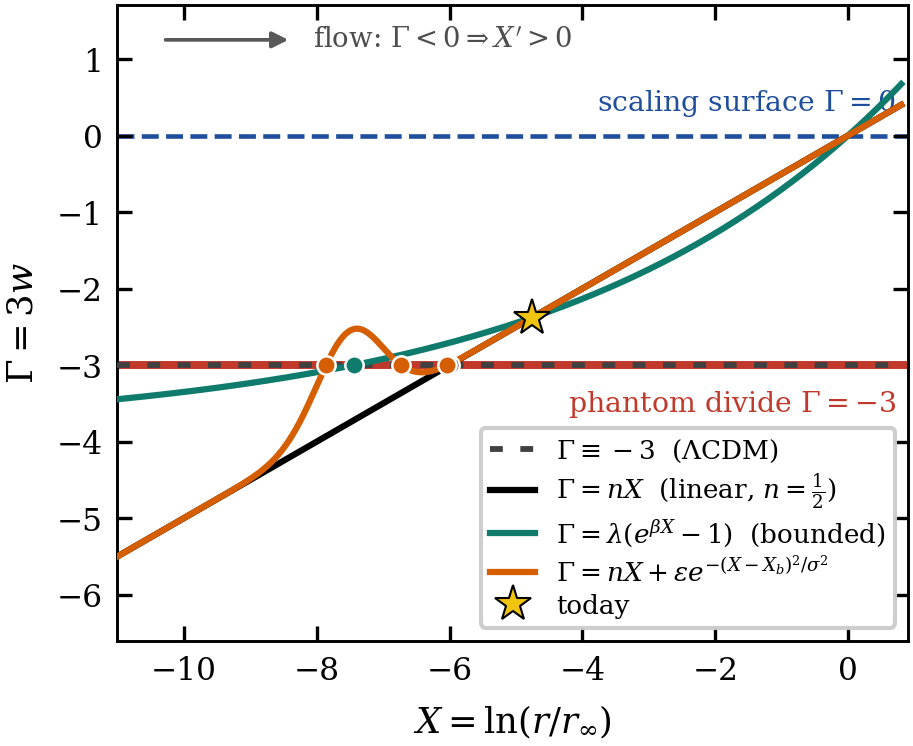}
\caption{\small The composition-force plane  and some examples of function $\Gamma(X)$.  The phantom divide is the fixed
horizontal line $\Gamma=-3$ (red); crossings are its transversal intersections
with $\Gamma(X)$ (dots) and their direction is the sign of the slope there.
Trajectories flow to the right wherever $\Gamma<0$, from $X\to-\infty$
($\rDE\ll\rho_m$) towards a stable zero at $X=0$ (blue dashed), which is a
dust-like scaling state.  $\LCDM$ is the constant force $\Gamma\equiv-3$: it
lies \emph{along} the divide at every epoch instead of crossing it.  The linear
force $\Gamma=nX$ is the family $w=w_0a^{-n}$, and its $n=1/2$ member is the
square-root model of Eq.~\eqref{eq:rPowerA}; the bounded force approaches $w\to-\lambda/3$ in the past
rather than diverging; the bump force crosses three times with alternating
direction.  The forces shown here have $\Gamma<-3$ as $X\to-\infty$ and therefore an
odd number of crossings.  The star marks
the present epoch, common to all curves
($\Omega_{m0}=0.32$, $w_0=-0.793$).}
\label{fig:forceplane}
\end{figure}

Wherever $w<0$, $X$ increases monotonically and can itself label successive
stages of the evolution. Along such a negative-pressure branch for DE we can {\it always} invert the history to
write $a=a(X)$ and introduce
\begin{equation}
 X'=-\Gamma(X),
 \qquad
 \Gamma(X)\equiv3w[a(X)].
 \label{eq:Gammaflow}
\end{equation}
The graph of $\Gamma$ then describes how rapidly the composition changes per
e-fold of expansion at each stage of the history. Some representative examples
for different types of evolution are shown in Figure \ref{fig:forceplane} (its  caption
can be better understood after reading the rest of this Section).
On the negative-pressure
branch, $\Gamma<0$ and the motion is towards increasing $X$. The autonomous
form of the equation follows from using a monotonic state variable as
a global clock controlling the dark sector evolution. So far, it does {\it not}
assert that a general microscopic theory has an equation of state depending
fundamentally on $X$ alone -- more on this in Section \ref{sec_cosflu}. Different trajectories of a theory with additional
degrees of freedom can generate different functions $\Gamma$. Once a flow is
specified as a model, however, its shape provides a compact way to connect the
observed evolution to its possible endpoints. See also
\cite{CopelandR,BahamondeR} for standard reviews for dynamical system applications
to dark energy models.

For a history approaching a finite, nonzero asymptotic ratio $\rinf$ (recall
e.g. the concrete example around Eq.\eqref{eq:rPowerA} in the
Introduction), the
physically natural choice, which we are going to adopt,  is $r_{\rm ref}=\rinf$, giving
\begin{equation}
 X=\ln \left({r}/{\rinf}\right).
 \label{eq:Xrinf}
\end{equation}
The final composition then lies at $X=0$, and $X$ measures the departure from
it. This choice identifies an endpoint when one exists. 

Let us make a couple of  examples:
\begin{itemize}
\item 
{\bf   $w=-1$, $\Gamma=$} constant. For the case
of $\Lambda$CDM we have
\begin{align}
 &\rho_\Lambda=\mathrm{const},
 \qquad
 r\propto a^3,
 \nonumber
 \\
 &X=3\ln a+\mathrm{const},
\end{align}
where $X$ is defined using a finite reference ratio. The corresponding flow is
\begin{equation}
 \Gamma_{\Lambda{\rm CDM}}\equiv-3.
 \label{eq:LCDMGamma}
\end{equation}
Thus $\LCDM$ can be represented by a horizontal line with no finite fixed point. See
Fig.~\ref{fig:forceplane}.
Although its dark-energy density is constant, its composition never becomes
stationary. Matter continues to dilute, and the ratio $r$ grows without bound
as the spacetime approaches de Sitter.
\item {\bf{  $w\propto a^{-1/2}$, $\Gamma= X/2$}}. 
Instead,  the example discussed
in the Introduction, Eq.~\eqref{eq:rPowerA}, corresponds to a simple linear function $\Gamma(X)  = X/2$.  For this choice of $\Gamma$, 
the evolution has a compositional  equilibrium point at $X=0$, with $r_\infty$ finite,
reproducing through very simple manipulations the results in formulas \eqref{eq:rPowerA} of the Introduction. 
We will return to further study and develop this class of models in Section \ref{sec:relaxation-model}.
\end{itemize}

\smallskip

A fixed composition and a constant
DE density are therefore physically different notions of a late-time state.
In what follows, inspired by the explicit example
in \cite{Borghetto2026}, we {\it assume} that 
a finite $r_\infty$ exists for  evolving dark energy 
scenarios --  hence with our normalization \eqref{eq:Xdef} for the composition variable relaxes to 
a compositional equilibrium at $X=0$ at late times. For the rest of this 
section we study  properties of trajectories heading towards
equilibrium, while in  Section \ref{sec:relaxation-model} we analyze 
a physically well motivated
choice for the relaxation function $\Gamma$.

\smallskip

Notice that the composition variable formalism is also useful beyond the two-component approximation of dark energy and pressureless matter.
Additional separately conserved species, such as radiation or neutrinos, change
$H$ and the total fractional DE density, but leave Eq.~\eqref{eq:Xidentity}
unchanged as long as the matter used in $r$ remains pressureless and separately
conserved. The relations between composition and total expansion must then
include these additional species. In what follows we use the two-component
system to make that relation explicit.

\subsection{Phantom crossing, density extrema and the number of crossings.}
\label{sec:crossing-geometry}

One  advantage of working with the composition variable $X$ is 
that it makes simpler to understand the global properties of the universe
evolution. For example,
phantom crossing concerns the change from a dark component whose density grows
as the universe expands to one whose density decreases, or the reverse. In
composition space, the corresponding equation of state $w=-1$ is the level
 $\Gamma(X_c)=-3$.
 The cosmological trajectory crosses the phantom divide when its flow curve
passes through this level. There is no obstruction to the composition clock
at this event: $X'=3$ is finite and positive. Although DE density evolution has an extremum at $w=-1$  (see Eq.~\eqref{eq:master}), $\rho_m$ is unaffected, making the evolution of
the ratio $\rho_{\rm DE}/\rho_m$ regular at the crossing.   Phantom crossing is therefore a
regular event in this background description, provided the flow itself is
regular. Whether a microscopic realization also has regular perturbations is
a separate question, which we address
in Section \ref{sec_cosflu}.

The local slope of the function $\Gamma(X)$ tells us which way the crossing proceeds. Differentiating
along the trajectory, and writing $\Gamma_X\equiv\dd\Gamma/\dd X$, gives
\begin{equation}
 w'=\frac{\Gamma_X X'}{3}
    =-\frac{\Gamma\Gamma_X}{3}.
\end{equation}
Consequently,
\begin{equation}
-\frac{\dd\ln|w|}{\dd\ln a}=\Gamma_X(X)\equiv n_{\rm eff}(z)
 \label{eq:crossingDirection}
\end{equation}
since $\ln|w|=\ln|\Gamma|-\ln 3$ and $X'=-\Gamma$, exactly, for any flow. The local slope of $\ln|w|$ against $\ln a$ {\it is} therefore the local relaxation rate for the system heading towards equilibrium. We can -- and we will -- use this property to test our scenarios.
 
At  transverse (i.e. not osculating) intersections, a positive slope describes an exit from the
phantom regime, whereas a negative slope describes entry into it. If the slope
vanishes, instead, higher-order behaviour must be examined to distinguish a genuine
crossing from a tangency. Hence, merely reaching $w=-1$ need not constitute a crossing.

The same geometry explains the extrema of the dark-energy density. Conservation
implies
\begin{equation}
 (\ln\rDE)'=-(3+\Gamma),
 \qquad
 (\ln\rDE)''=\Gamma\,\Gamma_X.
 \label{eq:rhoGamma}
\end{equation}
Below the level $\Gamma=-3$, the DE density grows; above it, the density falls.
At the crossing,
\begin{equation}
 (\ln\rDE)''_c=-3\Gamma_X(X_c).
\end{equation}
An upward crossing is therefore a local maximum of $\rDE$, and a downward
crossing is a local minimum. In particular, the upward crossing in a pulse-like
history marks the point at which DE stops growing in absolute density. Its
relative abundance can nevertheless continue to increase afterwards, as long
as $w$ remains negative. The peak of $\rDE$ is not a peak of $r$. See also 
\cite{Perkovic:2018pzv}.

Interestingly, 
this representation also makes the number and ordering of phantom crossings easy to
understand geometrically. On a connected interval with $\Gamma<0$, the universe moves through
each value of $X$ only once. Crossings in time correspond to sign-changing
intersections of $\Gamma(X)$ with the horizontal line $\Gamma=-3$ on the part
of the curve actually traversed. When all intersections are transverse, their
directions alternate: an exit from the phantom regime must be followed by an
entry before another exit is possible. Repeated phantom crossings thus require
structure in the flow curve, even though
-- as stated at the beginning -- the composition variable $X$ remains
monotonic.

\subsection{Acceleration and the asymptotic fate of the universe}
\label{sec:global-future}
 A negative-pressure component affects the
geometry through both its equation of state and its share of the energy
budget. For the matter--DE system, the total equation of state is
\begin{equation}
 w_{\rm tot}=w\Ode=\frac{\Gamma\Ode}{3}.
 \label{eq:wtotGamma}
\end{equation}
Thus component phantom behaviour, violation of the total null energy condition
(NEC), and accelerated expansion are three different statements about the composition
of our universe. The total
NEC, $\rho_{\rm tot}+p_{\rm tot}\ge0$, requires
\begin{equation}
 1+\frac{\Gamma\Ode}{3}\ge0,
\end{equation}
whereas accelerated expansion requires
\begin{equation}
 1+\Gamma\Ode<0.
\end{equation}
Acceleration is compatible with the total NEC in the interval
\begin{equation}
 -\frac{3}{\Ode}\le\Gamma<-\frac{1}{\Ode}.
 \label{eq:accelerationStrip}
\end{equation}
The positive matter contribution can therefore keep the total NEC satisfied
even when the effective DE component has $w<-1$ and violates it. Likewise, the onset or end
of acceleration is determined by $w\Ode=-1/3$, not by phantom crossing.
At $w=-1$, for example, acceleration occurs only if $\Ode>1/3$. A crossing
can take place while DE is still too subdominant to accelerate the universe.


The future destination of a composition history is controlled by the zeros
of its flow. A point satisfying
\begin{equation}
 \Gamma(X_\star)=0
 \label{cond_zer}
\end{equation}
is a stationary solution: the ratio of DE to matter no longer changes. For a
smooth flow, a small homogeneous displacement obeys, to linear order
around $X_\star$,
\begin{equation}
 \delta X'=-\Gamma_X(X_\star)\,\delta X.
\end{equation}
A positive slope makes the fixed point locally attractive. Physically, the
flow brings a composition slightly below the stationary ratio upwards, and a
composition slightly above it downwards. This tests stability within the
specified background flow. However, we stress that it does {\it not} establish the stability of all degrees
of freedom in a microscopic realization.

For a regular trajectory approaching  a point
as \eqref{cond_zer}, choosing the normalization
in Eq.~\eqref{eq:Xrinf} places it at $X=0$ and yields
\begin{equation}
 r\to\rinf,
 \qquad
 \Ode\to\frac{\rinf}{1+\rinf},
 \qquad
 w\to0,
 \qquad
 w_{\rm tot}\to0.
 \label{eq:scalingFuture}
\end{equation}
This is an equilibrium of the composition --  not a static universe. Both energy
densities continue to dilute, but their dilution rates become equal. Indeed,
since $\rho_m\propto a^{-3}$ and $r$ tends to a finite constant, the total
density asymptotically scales as $a^{-3}$.
Even if $\rinf>1$ and the component labelled DE carries most of the energy,
its pressure no longer sustains acceleration. Pressureless background behaviour
also does not, by itself, imply that this component clusters like cold dark
matter; that question belongs to  perturbation theory, which we
elaborate in Section \ref{sec_cosflu}.

\subsection{Global history of the universe.}
Any accelerated epoch on this branch of solutions must consequently end. The scaling future
has no big-rip singularity, and it also changes the long-term causal structure.
A future event horizon would require a finite remaining conformal time, whereas a pressureless matter-dominated scaling
 gives
$ \int^{\infty}{\dd t}/{a(t)}=\infty
$. There is therefore no future cosmological event horizon for a spacetime with
this regular asymptotic evolution. An apparent horizon still exists, though --  unlike an
event horizon, it is defined by the instantaneous expansion rather than the
entire future history. The stable scaling endpoint thus connects a finite
composition, the end of acceleration, and an unbounded future conformal time.
These conclusions follow from the assumed two-component, spatially flat
background and its approach to the fixed point. Additional components that
later dominate would require extra analysis.

\smallskip
To describe a global history, one must also check how far the chosen flow can
be continued towards the past. The condition $X\to-\infty$ means $r\to0$. But it 
does not  tell us whether this limit occurs only as $a\to0$ or already at a
finite scale factor. Along the negative-pressure branch,
$ \dd\ln a=-{\dd X}/{\Gamma(X)}$.
Hence the early boundary lies at infinite negative e-fold time precisely when
\begin{equation}
 \int_{-\infty}^{X_{\rm b}}\frac{\dd X}{|\Gamma(X)|}=\infty,
 \label{eq:pastExtent}
\end{equation}
where $X_{\rm b}$ is any finite interior point of the branch with $\Gamma<0$.
For an asymptotic flow $\Gamma\sim-C|X|^p$, with $C>0$, the integral diverges
for $p\le1$ and converges for $p>1$. For $p>1$
the background history is incomplete in the past, 
making the scenario much less appealing. 
The linear model $p=1$
has an interesting marginal behaviour -- it combines extension to $a\to0$ with a
stable future scaling state with finite $r_\infty$, providing a simple global realization of the
physical picture developed here. We study it next. 

\section{Linear relaxation for the dark sector}
\label{sec:relaxation-model}

The composition description developed in 
the previous Section allows us to formulate a
simple physical hypothesis. The dark sector composed
of dark energy and pressureless matter evolves towards a stationary
relative abundance, 
with the departure from that state relaxing at a
given fractional rate per e-fold of expansion. We now 
elaborate on this viewpoint, and
 select a specific form
for the function $\Gamma(X)$ in Eq.~\eqref{eq:Gammaflow}. The resulting 
setup is general and model independent, and it
assigns a physical interpretation to a well defined  dark-energy equation of state,
while providing a simple, physically motivated global history that can be tested against
observations. Moreover, as we will learn in Section \ref{sec_cosflu}, the same principles developed here
are instrumental to develop an understanding of the
dynamics of cosmic fluctuations. 

As done earlier, we choose the composition variable
$X=\ln(r/r_\infty)$ so that equilibrium lies at $X=0$, and write
$N=\ln a$. For a smooth flow with a stable fixed point,
the evolution near equilibrium necessarily takes the form
\begin{equation}
\begin{gathered}
 X'=-\Gamma(X),\qquad \Gamma(X)=nX+\mathcal{O}(X^2),\\
 n\equiv\Gamma_{,X}(0)>0.
\end{gathered}
 \label{eq:near_equilibrium_flow}
\end{equation}
To leading order, the displacement therefore decays exponentially
in $N$: over an interval $\Delta N$, its magnitude is reduced by
$e^{-n\Delta N}$. The parameter $n$ measures the relaxation rate
relative to cosmological expansion; in cosmic time, the corresponding
instantaneous rate is $nH$.

Our additional assumption is that this linear relaxation law remains
valid beyond the immediate neighbourhood of equilibrium.
A precise way to state the assumption is to require that rescaling
the displacement does not change its fractional relaxation rate.
If $X(N)$ is a solution, then $\lambda X(N)$ should satisfy the same
flow equation, with the same parameters and the same e-fold clock,
for constant $\lambda>0$ whenever both trajectories lie in the
domain of the model. This requires
$\Gamma(\lambda X)=\lambda\Gamma(X)$, 
and hence selects
\begin{equation}
\Gamma(X)=nX,\qquad n>0.
\label{eq_linans}
\end{equation}
Physically, the rate of change is proportional to the remaining
composition displacement: larger departures relax more rapidly
in absolute magnitude, but the time in e-folds needed to remove
a given fraction of the displacement is the same.
There is no additional characteristic displacement at which the
relaxation law changes. 

The connection with linear regression  provides a complementary
interpretation. Consider two neighbouring homogeneous solutions
of the same prescribed flow. Their separation obeys
\begin{equation}
\delta X'=-\Gamma_{,X}(X)\,\delta X.
\label{eq:neighbouring_flow}
\end{equation}
The fractional decay rates of the background displacement and
of this separation are therefore
\begin{equation}
-\frac{\mathrm d\ln|X|}{\mathrm dN}
=\frac{\Gamma(X)}{X},
\qquad
-\frac{\mathrm d\ln|\delta X|}{\mathrm dN}
=\Gamma_{,X}(X).
\label{eq:background_separation_rates}
\end{equation}
For the linear law, both quantities equal $n$. Conversely, demanding their
equality throughout a connected branch gives
$X\Gamma_{,X}=\Gamma$ and recovers Eq.~\eqref{eq_linans}.
Thus the rescaling property and equality of these two homogeneous
decay rates characterize the same background hypothesis.

Near equilibrium, the common relaxation rate is suggestive of
Onsager's regression hypothesis, which equates the relaxation rate
of spontaneous fluctuations to the macroscopic response to
small departures from equilibrium, induced by external agents \cite{Onsager1R,Onsager2R,deGrootMazurR}. Other works discussing non-equilibrium
thermodynamics in the context of gravity and cosmology
include \cite{IsraelStewartR,MaartensCausalR,ZimdahlBulkR,PavonWangR,WangGongAbdallaR,NonlinearISUnifiedR,BarotropicISR}.
A separation between background solutions is not, however,
automatically a physical cosmological fluctuation.
The latter requires a prescription for the pressure response,
velocities and metric perturbations. We therefore use
near-equilibrium regression as a guiding criterion for the
perturbative completion, rather than assume that it determines
the full fluctuation dynamics.
We anticipate that the relevant gauge-invariant composition
perturbation is
$\mathcal X=\delta X-w\,\delta_m$; its evolution, and the
conditions under which it recovers the background relaxation
rate, will be examined in   Section \ref{sec_cosflu}.

The linear profile for the function $\Gamma(X)$ has an immediate
consequence for the background evolution.
Combining $X'=-nX$ with the exact identity
$X'=-3w$, and normalizing $a=1$ today, gives
\begin{equation}
X(a)=X_0a^{-n},
\qquad
w(a)=w_0a^{-n},
\qquad
w_0={nX_0}/{3}.
\label{eq:linear_relaxation_history}
\end{equation}
The power-law equation of state is therefore the cosmological
history generated by constant fractional relaxation of the
logarithmic composition. Its two parameters have distinct
roles: $n$ specifies the relaxation law, while $w_0$ locates
the present state along its trajectory. This 
case generalizes the example of Eq.~\eqref{eq:rPowerA}. Related discussions
and 
parametrizations for the DE equation of state
can be found e.g. in \cite{MaZhangR,FengR,LiZhangDivR,BarbozaR,JassalR,RezaeiR,HussainR,LeeShapeR,DeFeliceFastR,CorasanitiR}.
We next develop the  consequences of this hypothesis at the background level,
before confronting them with cosmological observations in Section \ref{sec:observables}.

\subsection{Consequences of the linear relaxation hypothesis.} 
  We adopt 
the linear law of relaxation \eqref{eq_linans}
as a mimimal global background hypothesis, and we examine the history it generates, from
the early suppression of dark energy to the end of accelerated expansion.
Our analysis generalizes the discussion
in \cite{Borghetto2026}.

\begin{figure*}[t]
\centering
\paperfigure{1.85in}{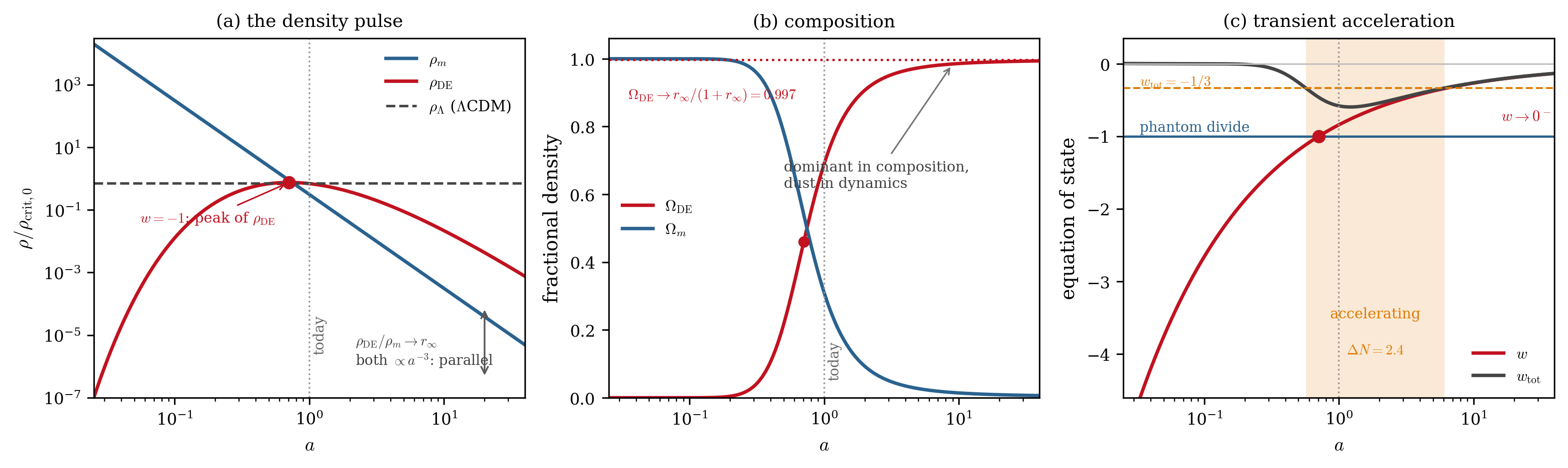}
\caption{\small The global history generated by a single linear relaxation flow,
$\Gamma(X)=nX$ with $n=1/2$, shown over four decades in scale factor for
$\Omega_{m0}=0.31$ and $w_0=-0.84$.
\emph{(a)} Densities in units of the present critical density. The dark
component rises from a non-perturbatively suppressed early branch, reaches a
maximum exactly at the phantom crossing (dot), and then dilutes. At late times
its slope becomes equal to that of matter: both scale as $a^{-3}$. A cosmological constant (dashed) never becomes
parallel to matter, and its ratio to matter grows without bound.
\emph{(b)} Fractional densities. The asymptotic state is dark-energy dominated
in composition, $\Omega_{\rm DE}\to r_\infty/(1+r_\infty)=0.997$, while being
pressureless in its dynamics.
\emph{(c)} Equation of state. The component crosses the phantom divide once,
upwards, while the total equation of state remains well above $-1$ throughout.
Acceleration occupies the shaded window of $\Delta N=2.4$ e-folds, bounded by
$w_{\rm tot}=-1/3$ rather than by the crossing, and ends at finite scale
factor as $w\to0^-$.}
\label{fig:globalhistory}
\end{figure*}

 With $a=1$ today, the solution of the evolution equation is
\begin{equation}
 X(a)=X_0a^{-n}.
 \label{eq:XlinearSolution}
\end{equation}
From now on,  $X_0=X(a=1)$ denotes the present displacement; the
fixed point is at $X=0$. Using $\Gamma=3w$, we obtain
\begin{equation}
 w(a)=w_0a^{-n},
 \qquad
 X_0=\frac{3w_0}{n}.
 \label{eq:wPowerLaw}
\end{equation}
This functional form has been constrained phenomenologically against DESI DR2 and supernovae by Hossain \cite{HossainR}, who finds that parametrisations permitting stronger phantom excursions fit better. We will specialize the discussion to our physical setup in Section \ref{sec:observables}. The constant-$\Gamma$ member of the family is the $\xi$ parametrisation of Dalal et al. \cite{DalalR}. What our composition description adds is a physical   interpretation of why this form should 
arise, and its physical consequences (including the perturbation dynamics). 
We concentrate on the negative-pressure branch $w_0<0$, for which $X_0<0$.
On this branch the composition approaches equilibrium from below. The ratio
of DE to matter increases, while the magnitude of the negative equation of
state decreases. See Fig.~\ref{fig:globalhistory}.

\paragraph*{Roles of the available parameters.} 

The different roles of the parameters involved are worth emphasizing. In the coordinate
centred on equilibrium, the straight line $\Gamma=nX$ has a single response
coefficient: its slope $n$. This coefficient specifies the rate at which a
composition displacement decays. The integration constant $X_0$ (or
equivalently $w_0$) specifies the present position on a solution of that law.
For a given $n$, different choices of $X_0$ describe universes at different
stages of relaxation today. Conversely, at fixed $w_0$, changing $n$ changes
both the rate and the present displacement needed to reproduce that equation
of state. The asymptotic ratio also has to be located in physical abundance
space: once the present ratio $r_0$ is specified, $\rinf$ is determined by
$X_0$. It is not an additional independent parameter.

There is consequently {\it one} parameter $(n)$
controlling the response law towards
equilibrium, but {\it two}
DE quantities, $(w_0,n)$ needed for a general cosmological fit -- in addition to the
usual background parameters such as the present-day matter abundance. Hence one quantity characterizes the evolution
law -- while the other fixes its present state during cosmic evolution. 

\paragraph*{Comparing with CPL.} 
For comparison, CPL writes
\begin{equation}
 w_{\rm CPL}(a)=w_0+w_a(1-a),
\end{equation}
with the present value and the slope in $a$ as its two coefficients. Near the
present epoch the two descriptions can be matched (see also \cite{Borghetto2026}). Expanding
Eq.~\eqref{eq:wPowerLaw},
\begin{equation}
 w(a)=w_0+n w_0(1-a)
       +\frac{n(n+1)}{2}w_0(1-a)^2
       +\mathcal O[(1-a)^3],
\end{equation}
so equality of the present value and first derivative gives
\begin{equation}
 w_a=nw_0.
 \label{eq:waMap}
\end{equation}
We emphasize that
this is a local matching relation, not an exact identification of the two
histories or of parameters inferred from fits over a finite redshift range.
For $w_0<0$ and $n>0$, it selects a negative CPL slope: the equation of state
was more negative in the recent past. At fixed $w_0$, the ratio of this local
slope to $w_0$ measures the relaxation rate.

The physical difference becomes clearer beyond first order. In the relaxation
model, the fractional change of $w$ per e-fold remains constant, giving
\begin{equation}
 \frac{\dd^k w}{\dd (\ln a)^k}=(-n)^k w,
 \qquad k=1,2,\ldots.
 \label{eq:derivativeHierarchy}
\end{equation}
Once the rate and the current state are known, no additional coefficients
are needed to specify the curvature or higher derivatives. 
The predicted futures illustrate the differences between CPL and linear
relaxation. For a locally matched CPL history
with $w_a=nw_0<0$, $w_{\rm CPL}$ eventually becomes positive and grows without
bound as $a\to\infty$. In the relaxation case, $w$ approaches zero from
below and the composition settles to a finite ratio. CPL remains useful as a
description over the observed redshift range, but its coefficients do not
encode such scaling endpoint. The additional physical commitment in the
relaxation model is precisely that the same rate linking nearby epochs also
governs the global approach to a specified kind of future. 

\subsection{The phantom divide and its implications.} 

The abundance and density histories follow from
$r=\rinf e^X$ and matter conservation:
\begin{align}
 r(a)
 &=r_0\exp\!\left[
 \frac{3w_0}{n}\left(a^{-n}-1\right)
 \right],
 \label{eq:rLinearExact}\\
 \rDE(a)
 &=\rho_{{\rm DE},0}\,a^{-3}\exp\!\left[
 \frac{3w_0}{n}\left(a^{-n}-1\right)
 \right].
 \label{eq:rhoLinearExact}
\end{align}
In particular,
\begin{equation}
 \rinf=r_0\exp\!\left(-\frac{3w_0}{n}\right).
 \label{eq:rinfLinear}
\end{equation}
Since $w_0<0$, the final ratio exceeds its present value. Dark energy continues
to gain in relative abundance even as its pressure becomes less negative.
For fixed present $r_0$ and $w_0$, a smaller rate implies a larger remaining
change in composition. The weaker the fractional relaxation per e-fold, the
longer the negative pressure continues to separate the dilution rates of the
two components.

For the branch $-1<w_0<0$, the universe today lies after phantom crossing.
Solving $w(a_c)=-1$ gives for the scale factor at crossing: 
\begin{equation}
 a_c=|w_0|^{1/n},
 \qquad
 X_c=-\frac{3}{n}.
 \label{eq:acLinear}
\end{equation}
The same expressions apply for all $w_0<0$: if $w_0<-1$ the crossing lies in
the future, and if $w_0=-1$ it occurs today. For the arguments developed in
Section \ref{sec:state-space}, because $w$ increases monotonically
from $-\infty$ to zero, there is exactly one crossing. Its direction is fixed
by the positive slope $\Gamma_X=n$, and it marks the unique maximum of the
absolute DE density. At fixed $n$, every trajectory on this branch crosses at
the same composition displacement $X_c$. The integration constant determines
when that point is reached. The timing of acceleration is different, because
it also depends on the abundance $r_0$ through the total equation of state. See Fig.~\ref{fig:globalhistory} for a representation of these behaviours. 

\subsection{Past and future histories.}
The early limit shows why this history can contain a strongly phantom
component without making it dynamically important in the early universe.
Writing $\rho_{m,0}$ for today's matter density, as $a\to0$ we have
\begin{equation}
 r(a)=\rinf\exp\!\left(\frac{3w_0}{n}a^{-n}\right)\to0,
 \label{eq:rEarly}
\end{equation}
and
\begin{equation}
 \rDE(a)=\rho_{m,0}\rinf\,a^{-3}
 \exp\!\left(\frac{3w_0}{n}a^{-n}\right)\to0.
 \label{eq:rhoEarly}
\end{equation}
The negative exponent dominates every inverse power of $a$. Thus both the
relative abundance and the absolute density vanish towards the early boundary
of the effective solution. Although $|w|\to\infty$, the pressure
$p_{\rm DE}=w\rDE$ also tends to zero. Its extra power of $a^{-n}$ can not
compensate for the exponential suppression. A large equation-of-state ratio
therefore does not imply a large stress-energy contribution. Notice that radiation can be added
to the early background without altering the separately conserved DE and
matter histories.

\medskip

In the opposite limit, $a\to\infty$,
\begin{equation}
 X\to0^-,
 \qquad
 w\to0^-,
 \qquad
 r\to\rinf,
\end{equation}
and hence
\begin{equation}
 \rDE\sim\rho_{m,0}\rinf\,a^{-3}.
\end{equation}
The late-time
composition is stationary even though both densities continue to fall.
 Any accelerated epoch is therefore transient, although
the existence of such an epoch is not guaranteed by $n>0$ and $w_0<0$ alone.
DE must become sufficiently abundant while its pressure is sufficiently
negative. For histories reproducing the observed acceleration, the model
predicts its eventual end.

One can also determine when the component phantom phase respects the total
null energy condition. In the two-component system,
\begin{equation}
 1+w_{\rm tot}
 =\frac{1+r(1+w)}{1+r}.
 \label{eq:linearNECgeneral}
\end{equation}

After phantom crossing, both terms in the numerator are nonnegative. The
possible competition occurs earlier, when $w<-1$: the negative contribution
$r(|w|-1)$ must not outweigh the positive matter contribution. The exact solution gives $r=\rinf e^{-3|w|/n}$. On the phantom
interval $|w|>1$, the quantity $r(|w|-1)$ reaches its maximum at
$|w|=1+n/3$. Requiring this maximum to be no greater than unity yields
\begin{equation}
 \frac{n r_0}{3}
 \exp\!\left[
 -\frac{3(1-|w_0|)}{n}-1
 \right]\le1.
 \label{eq:linearNECcriterion}
\end{equation}
For the full negative-pressure solution this is the condition for the total
NEC to hold at every finite $a>0$. The strongest test is at an intermediate
stage of the phantom phase, where DE has acquired a significant abundance
but still has sufficiently negative pressure. At very early times its
exponential suppression wins, while at crossing its contribution to
$\rho+p$ vanishes.  Notice that this does
{\it not} remove the usual early-time singularity of the matter or radiation
background, or settle the perturbative consistency of the effective DE model.

The square-root history that early on motivated our construction -- recall the discussion
around Eq.~\eqref{eq:rPowerA} in the Introduction --is the benchmark
$n=1/2$.
 The model then uses one fewer DE parameter
than CPL while retaining the early suppression, unique density maximum and
pressureless scaling future. See Fig.~\ref{fig:globalhistory}. Fixing $n=1/2$ defines a useful restricted model, which is in good agreement
with data (see \cite{Borghetto2026} and the next Section).

\section{Observational tests of linear relaxation}
\label{sec:observables}
\label{sec:observations}

In this Section
we test the linear relaxation hypothesis $\Gamma = n X$ of
Section \ref{sec:relaxation-model} against current data, we compare against
CPL, and we discuss
physical implications of our findings. Since we focus
only on background evolution, we concentrate  on cosmological
probes of geometrical aspects of our universe dynamics. 
Related studies of state-space diagnostics and null
tests can be found e.g. in \cite{CaldwellLinderR,SahniOmR,ShafielooR,StatefinderR,ZunckelR,StatefinderHierarchyR,ScherrerWWR,DindaR,OsanoR}. Further  analysis testing
  dark energy scenarios with recent data can be found \cite{DESIDR2A,DESIExtendedR,LodhaR,ChenHuangWangR,SchererR,GialamasR,LiZhangSignR,SilvaIDER,RiessR}.

\begin{table*}[t]
\centering
\renewcommand{\arraystretch}{1.22}
\begin{tabular}{lcccc}
\toprule
data & best-fit $n$ & $68\%$ interval & $\Delta\chi^2$ vs.\ $w$CDM
& nominal significance\\
\midrule
BAO $+$ CMB      & $0.803$ & $[0.47,1.20]$ & $7.02$  & $2.6\sigma$\\ %
$+$ Pantheon$^+$ & $0.383$ & $[0.24,0.54]$ & $7.86$  & $2.8\sigma$\\ %
$+$ DES-SN5YR    & $0.433$ & $[0.28,0.59]$ & $9.62$  & $3.1\sigma$\\ %
$+$ Union3       & $0.653$ & $[0.45,0.87]$ & $13.13$ & $3.6\sigma$\\ %
\bottomrule
\end{tabular}
\caption{\small Constraints on the relaxation rate. The comparison with $w$CDM
isolates the evidence for time evolution, since $n\to0$ is the constant-$w$
limit of the family. Best-fit values are the minimisers of the full
likelihood; the intervals are profile-likelihood bounds read off the grid of
Fig.~3. The quoted significances are the nominal $\sqrt{\Delta\chi^2}$ values
for one additional parameter.}
\label{tab:relaxationRate}
\end{table*}

\begin{figure*}[t]
\centering
\paperfigure{1.85in}{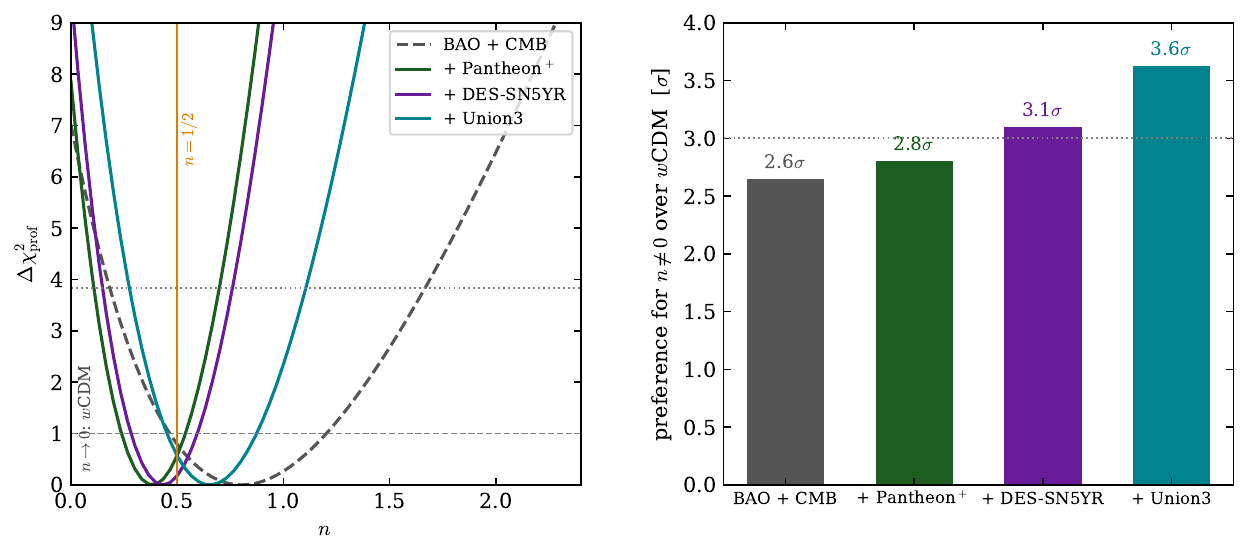}
\caption{\small \emph{Left:} profile likelihood in the relaxation rate for the
BAO+CMB baseline and for each supernova combination.  The $n\to0$ endpoint is
$w$CDM, while the vertical line marks $n=1/2$.
\emph{Right:} nominal preference for a non-zero relaxation rate relative to
$w$CDM.}
\label{fig:an-profiles}
\end{figure*}

\subsection{Data and likelihoods.}

Our baseline data set combines the DESI DR2 BAO likelihood with the Planck 2018
distance priors.  The DESI likelihood contains 13 measurements of
$D_V/r_d$, $D_M/r_d$ and $D_H/r_d$ over
$0.295\leq z\leq2.33$, with the published covariance
\cite{DESIDR2A}.  The CMB information is compressed into the distance-prior
variables $(R,\ell_A,\omega_b)$ of Ref.~\cite{ChenHuangWangR}.  We use distance
priors rather than the full Planck likelihood, so to preserve the
high-redshift geometric anchor, avoiding assumptions about the perturbations
of the effective dark components. As a check that this compression is adequate for the present purpose, our CPL
fit to BAO $+$ CMB alone gives
$(w_0,w_a,H_0,\Om)=(-0.437,-1.670,63.90,0.351)$, to be compared with
$(-0.45,-1.61,63.9,0.349)$ obtained from a full Boltzmann analysis of Planck
2018 $+$ DESI DR2 in VCDM \cite{AroraR}. The agreement of a compressed geometric likelihood
with a complete CMB treatment at this level indicates that the distance priors
retain the information relevant to late-time dark-energy constraints.

The low-redshift shape information is supplied by three supernova compilations,
which we add \emph{separately} to the BAO+CMB baseline:
Pantheon$^+$~\cite{BroutA}, DES-SN5YR~\cite{DESSNA}, and
Union3~\cite{RubinA}.  For Pantheon$^+$ we use the 1580 objects surviving the
standard Hubble-diagram cut after removing the Cepheid calibrators.  For DES we
use the Dovekie photometric recalibration currently distributed with the public
data \cite{Popovic:2025glk,DES:2025sig}; this distinction matters for
direct comparison with published DES-Y5 parameter values.  Union3 is used in its
22-bin compressed form. Further details on our numerical procedure can be 
found in Appendix \ref{app:method}. 

For each supernova sample the unknown absolute magnitude is analytically
marginalized.  If $\bm r$ denotes the vector of distance-modulus residuals,
\begin{equation}
\begin{gathered}
 \chi^2_{\rm SN}=A-\frac{B^2}{C},\\
 A=\bm r^{\mathsf T}\bm C^{-1}\bm r,\qquad
 B=\bm r^{\mathsf T}\bm C^{-1}\bm 1,\\
 C=\bm 1^{\mathsf T}\bm C^{-1}\bm 1.
\end{gathered}
 \label{eq:SNmarg}
\end{equation}
The supernovae therefore constrain the shape of the distance--redshift relation,
but do not by themselves determine $H_0$.  This is  complementary to the
BAO+CMB combination: the latter supplies the acoustic calibration and high-redshift
lever arm, while the supernovae densely sample the region below $z\simeq0.3$ where
the present equation of state and its local evolution are best constrained.

We compare five cosmological  histories on the same data described above:
\begin{align}
 \Lambda{\rm CDM}:&
 \qquad w=-1,
 \nonumber\\
 w{\rm CDM}:&
 \qquad w=w_0,
 \nonumber\\
 \text{linear relaxation}:&
 \qquad w=w_0a^{-n},
 \nonumber\\
 n=\tfrac12\ \text{relaxation}:&
 \qquad w=w_0a^{-1/2},
 \nonumber\\
 {\rm CPL}:&
 \qquad w=w_0+w_a(1-a).
 \label{eq:modelsCompared}
\end{align}
The common background parameters are $(\Omega_m,H_0,\omega_b)$.  We quote
one-dimensional profile-likelihood intervals, using
$\Delta\chi^2=1$ and $3.84$ for the nominal $68\%$ and $95\%$ intervals.

\subsection{Measuring the relaxation rate.}

The constraint on the relaxation rate is summarized in
Table~\ref{tab:relaxationRate}.  The most important comparison is with $w$CDM,
rather than directly with $\Lambda$CDM.  The physics
addressed by the quantity  $n$ is whether
the dark component evolves away from a constant equation of state, and
$n\rightarrow0$ is exactly the constant-$w$ endpoint of the relaxation family.

The effect of the supernovae is important.  With BAO and compressed CMB geometry
alone, $w_0$ and $n$ are strongly correlated. The data constrain an integrated
dark-energy history over a long redshift interval, but have comparatively little
leverage on its local slope near the present epoch.  The dense low-redshift
supernova sampling cuts across this degeneracy, reducing the uncertainty on $n$
from approximately $0.4$ to approximately $0.15$.

Equally important is the stability of the result
when changing data sets.  The three supernova
compilations give best-fit rates $n=0.38$, $0.43$ and $0.65$, respectively. 
All three are consistent with $n=1/2$ at $68\%$.  The central values are not
identical --  nor should they be expected to be given the differences among the
supernova samples -- but they select the same range of relaxation timescales. The
preferred values for $n$
correspond to a composition mode relaxing towards
equilibrium on a timescale of order the Hubble
time.

Figure~\ref{fig:an-profiles} shows the profile likelihoods explicitly.  The
$n\to0$ endpoint is the $w$CDM model, while the vertical line marks the
square-root rate $n=1/2$.  The preference for a non-zero rate strengthens when
supernovae are included.

\begin{figure*}[t]
\centering
\paperfigure{1.85in}{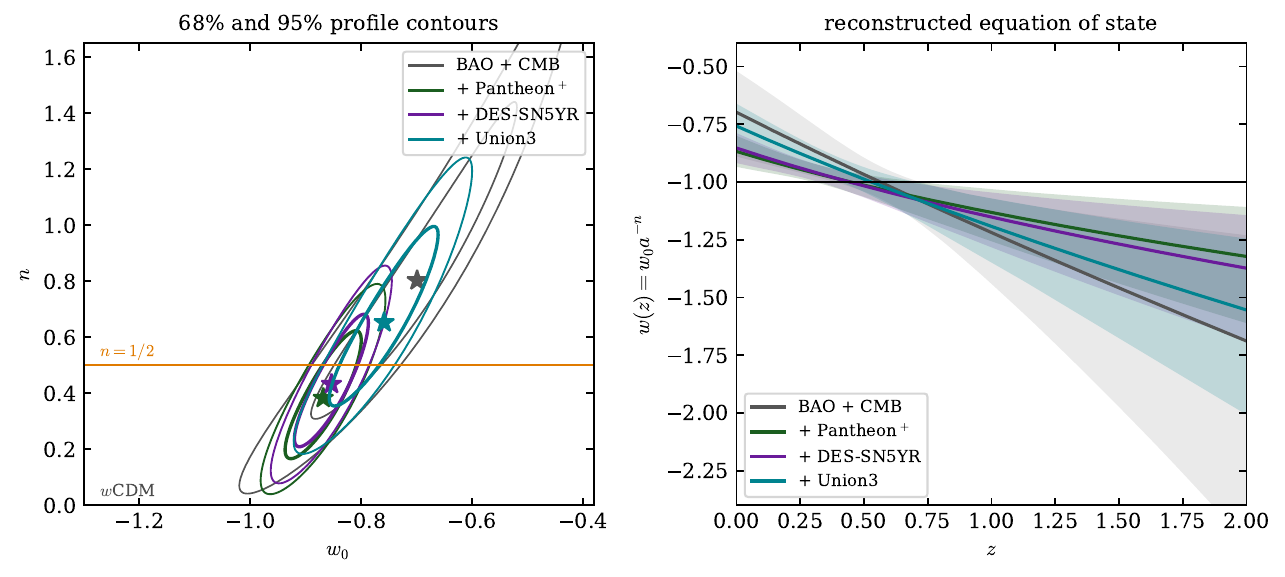}
\caption{\small \emph{Left:} two-dimensional profile-likelihood contours in
$(w_0,n)$.  Supernovae break the geometry-only degeneracy between the present
equation of state and the relaxation rate.
\emph{Right:} corresponding reconstruction of $w(a)$; the three
supernova-inclusive fits agree closely and are all consistent with the
$n=1/2$ trajectory.}
\label{fig:an-contours}
\end{figure*}

The origin of the improvement is visible in Fig.~\ref{fig:an-contours}.  BAO plus
CMB geometry allows an extended degeneracy in the $(w_0,n)$ plane; the
supernovae constrain the low-redshift shape and intersect this direction almost
orthogonally.  The resulting reconstructions of $w(a)$ are  more tightly
clustered than the geometry-only solution.

\subsection{Model comparison and the role of the square-root example.}

The full model comparison is given in Table~\ref{tab:modelComparison}.  We quote
the Akaike information criterion,
\begin{equation}
 {\rm AIC}=\chi^2_{\rm min}+2k_{\rm par},
\end{equation}
with $k_{\rm par}$ the number of parameters, 
and report $\Delta{\rm AIC}$ relative to the best model for each data
combination.

\begin{table*}[t]
\centering
\small
\renewcommand{\arraystretch}{1.16}
\begin{tabular}{llcccrr}
\toprule
data & model & $\Omega_m$ & $H_0$ & $w_0$ & $\chi^2_{\rm min}$ & $\Delta{\rm AIC}$\\
\midrule
\multirow{5}{*}{BAO $+$ CMB}
 & $\Lambda$CDM        & $0.3000$ & $68.64$ & $-1$     & $15.402$ & $4.68$\\
 & $w$CDM              & $0.2957$ & $69.25$ & $-1.026$ & $14.951$ & $6.23$\\
 & relaxation          & $0.3322$ & $65.72$ & $-0.699$ & $7.934$  & $1.21$\\
 & $n=1/2$             & $0.3198$ & $66.84$ & $-0.805$ & $8.726$  & $0.00$\\
 & CPL                 & $0.3513$ & $63.90$ & $-0.437$ & $7.310$  & $0.58$\\
\midrule
\multirow{5}{*}{$+$ Pantheon$^+$}
 & $\Lambda$CDM        & $0.3015$ & $68.53$ & $-1$     & $1404.967$ & $6.28$\\
 & $w$CDM              & $0.3049$ & $68.03$ & $-0.976$ & $1403.977$ & $7.29$\\
 & relaxation          & $0.3113$ & $67.75$ & $-0.868$ & $1396.114$ & $1.43$\\
 & $n=1/2$             & $0.3130$ & $67.68$ & $-0.839$ & $1396.686$ & $0.00$\\
 & CPL                 & $0.3109$ & $67.73$ & $-0.849$ & $1397.072$ & $2.39$\\
\midrule
\multirow{5}{*}{$+$ DES-SN5YR}
 & $\Lambda$CDM        & $0.3019$ & $68.50$ & $-1$     & $1650.648$ & $9.11$\\
 & $w$CDM              & $0.3058$ & $67.91$ & $-0.971$ & $1648.975$ & $9.44$\\
 & relaxation          & $0.3124$ & $67.67$ & $-0.853$ & $1639.356$ & $1.82$\\
 & $n=1/2$             & $0.3133$ & $67.64$ & $-0.837$ & $1639.534$ & $0.00$\\
 & CPL                 & $0.3127$ & $67.59$ & $-0.819$ & $1639.741$ & $2.21$\\
\midrule
\multirow{5}{*}{$+$ Union3}
 & $\Lambda$CDM        & $0.3012$ & $68.55$ & $-1$     & $43.924$ & $11.44$\\
 & $w$CDM              & $0.3055$ & $67.95$ & $-0.973$ & $43.050$ & $12.57$\\
 & relaxation          & $0.3243$ & $66.48$ & $-0.758$ & $29.917$ & $1.43$\\
 & $n=1/2$             & $0.3203$ & $66.78$ & $-0.803$ & $30.484$ & $0.00$\\
 & CPL                 & $0.3264$ & $66.20$ & $-0.684$ & $30.026$ & $1.54$\\
\bottomrule
\end{tabular}
\caption{\small Background model comparison among
the possibilities of Eq.~\eqref{eq:modelsCompared}.  $H_0$ is in
${\rm km\,s^{-1}\,Mpc^{-1}}$.  CPL has a second equation-of-state parameter
$w_a=-1.670,-0.531,-0.644,-0.998$ for the four data combinations, 
respectively.  The fixed $n=1/2$ relaxation history has the lowest AIC in every
combination considered.}
\label{tab:modelComparison}
\end{table*}

Two aspects of this table are worth stressing.  The general relaxation model and
CPL contain the same number of fitted background parameters, and their minimum
$\chi^2$ values are very close.  Present distance data therefore do not select
between the two functional forms on goodness of fit alone.  However,
the motivation for our relaxation description is {\it not} that it can
produce a dramatically lower $\chi^2$ than CPL. What is important are the physical
motivations and interpretation of the results. 

An  interesting comparison is  with the fixed $n=1/2$ trajectory.
Once the relaxation rate is fixed, the only dark-energy parameter left in the
background solution is the integration constant $w_0$.  Nevertheless this
one-parameter dark-energy history matches the two-parameter CPL fit to within
$\Delta\chi^2\lesssim1.5$ in every data combination.  The resulting AIC
differences relative to CPL are $0.58$, $2.39$, $2.21$ and $1.54$ in favour of 
the square-root model.  These are modest differences and should not be described
as decisive model selection.  Their significance is instead one of economy.
The same quality of fit is obtained after replacing an independent Taylor
coefficient by a fixed dynamical rate. This is in overall agreement
with the findings of \cite{Borghetto2026}.  In Section \ref{sec_cosflu}
we also show that the special case $n=1/2$ has interesting
properties for the evolution of perturbations.

This result is closely connected with the distinction made in
Sec.~\ref{sec:relaxation-model}.  In CPL, $(w_0,w_a)$ specify independently the
value and local slope of the trajectory.  In the relaxation model the slope is
generated by the evolution law, while $w_0$ is an integration constant specifying
the present state.  The fact that $n=1/2$ remains compatible with all three
supernova compilations therefore makes the square-root model a useful benchmark,
even though present data do not establish that this particular rate is fundamental.

\paragraph*{What is actually distinguished from CPL?}
\label{sec:data-cpl}

The comparison with CPL changes appreciably once supernovae are included.  On
BAO+CMB geometry alone, the CPL best fit lies at $w_0=-0.437$.  Supernovae drive the CPL present-day equation of state to
$w_0=-0.85$, $-0.82$ and $-0.68$ for the three compilations, much closer to the
corresponding relaxation values.  

 Within the redshift interval populated by supernovae, CPL and
relaxation can produce nearly indistinguishable Hubble diagrams.  Their main
difference is therefore not a presently resolved feature in the distance data,
but the global structure our our universe's evolution that accompanies the fit.  
The relaxation family has non-perturbatively suppressed early dark energy and
a fixed late-time scaling state with $w\to0$, and generates both from a single
evolution law. CPL is a local expansion and is not designed to provide
corresponding asymptotic information. We stress that the density maximum at
the phantom crossing is \emph{not} among the distinguishing features: by
Eq.~(II.12) the relation $\rDE'=0\Leftrightarrow w=-1$ holds for any separately
conserved component, so a CPL history with $w_0+w_a<-1$ also possesses a
phantom past, an upward crossing and a density pulse. What differs is the
early-time suppression, which is exponential here and power-law for CPL, and
the late-time endpoint, which CPL does not possess.

One practical consequence is that the crossing epoch is likely to be a more
robust quantity than the individual Taylor coefficients of CPL.  
In the relaxation language it is a state-space event, easily calculable in terms of the available parameters. In
a generic parametrization, instead,  the same event can be represented by rather
different pairs of coefficients. It should be emphasized that, within the
relaxation family, $a_c=|w_0|^{1/n}$ is an identity satisfied by construction
rather than a test. It acquires
testable content only when confronted with an independent, model-agnostic
reconstruction of the crossing epoch and of $w_0$. What the relation does
provide immediately is an interpretation: it states that the measured $w_0$
fixes the position of the present epoch along the relaxation trajectory, since
$X_0/X_c=|w_0|$.

\paragraph*{Implications and limitations of our data analysis.}

The inclusion of supernovae also removes two apparent costs of the geometry-only
fit.  The BAO+CMB relaxation solution preferred
$H_0=65.7\,{\rm km\,s^{-1}\,Mpc^{-1}}$, substantially below the corresponding 
$\Lambda$CDM value.  With Pantheon$^+$ and DES-SN5YR the best-fit value moves to
approximately $67.7\,{\rm km\,s^{-1}\,Mpc^{-1}}$, while Union3 gives 
$66.5\,{\rm km\,s^{-1}\,Mpc^{-1}}$.  The relaxation model therefore neither 
resolves the Hubble tension nor, once supernovae are included, substantially
worsens it relative to $\Lambda$CDM on the same data.

We do not quote an $S_8$ prediction from this background analysis.  Estimating
$\sigma_8$ requires a perturbation treatment and a Boltzmann calculation; simple
transfer-function estimates are adequate for qualitative model-to-model trends
but not for a competitive absolute constraint. 
See \cite{AroraR,Borghetto2026} for an analysis within the (perturbatively-controlled) VCDM scenario. 

Several further qualifications are worth keeping in mind.  The CMB information
used here is compressed and therefore excludes lensing and ISW information.  The
Planck distance priors assume the standard early-time sector.  The DES sample is
the Dovekie recalibration, so the quoted numbers should not be compared directly
with the original DES-SN5YR cosmological constraints.  The three supernova
compilations also share some low-redshift objects and are therefore not
statistically independent; we use them as separate robustness tests rather than
combining them.  Finally, the quoted significances are profile-likelihood
quantities, not Bayes factors.

With these caveats in mind, the observational message is nevertheless
simple and physically clear.  Present
background data do not distinguish relaxation from CPL through a unique feature of
the Hubble diagram.  They do show that a non-zero relaxation rate is preferred
over constant $w$ in all data combinations considered; that the preferred rate is
stable under the addition of three different supernova compilations and remains
consistent with $n=1/2$; and that the fixed square-root trajectory reaches the
same fit quality as CPL with one fewer dark-energy fit parameter.  The sharper
tests of the framework are therefore the structural ones.
 In Appendix \ref{app_null} we make further tests of the linearity of
$\ln|w|$, finding results
in  support of our linear relaxation framework.

\subsection{When the DE density peaks and acceleration ends.}
\label{sec:observational-epochs}

\begin{figure*}[t]
\centering
\paperfigure{1.0in}{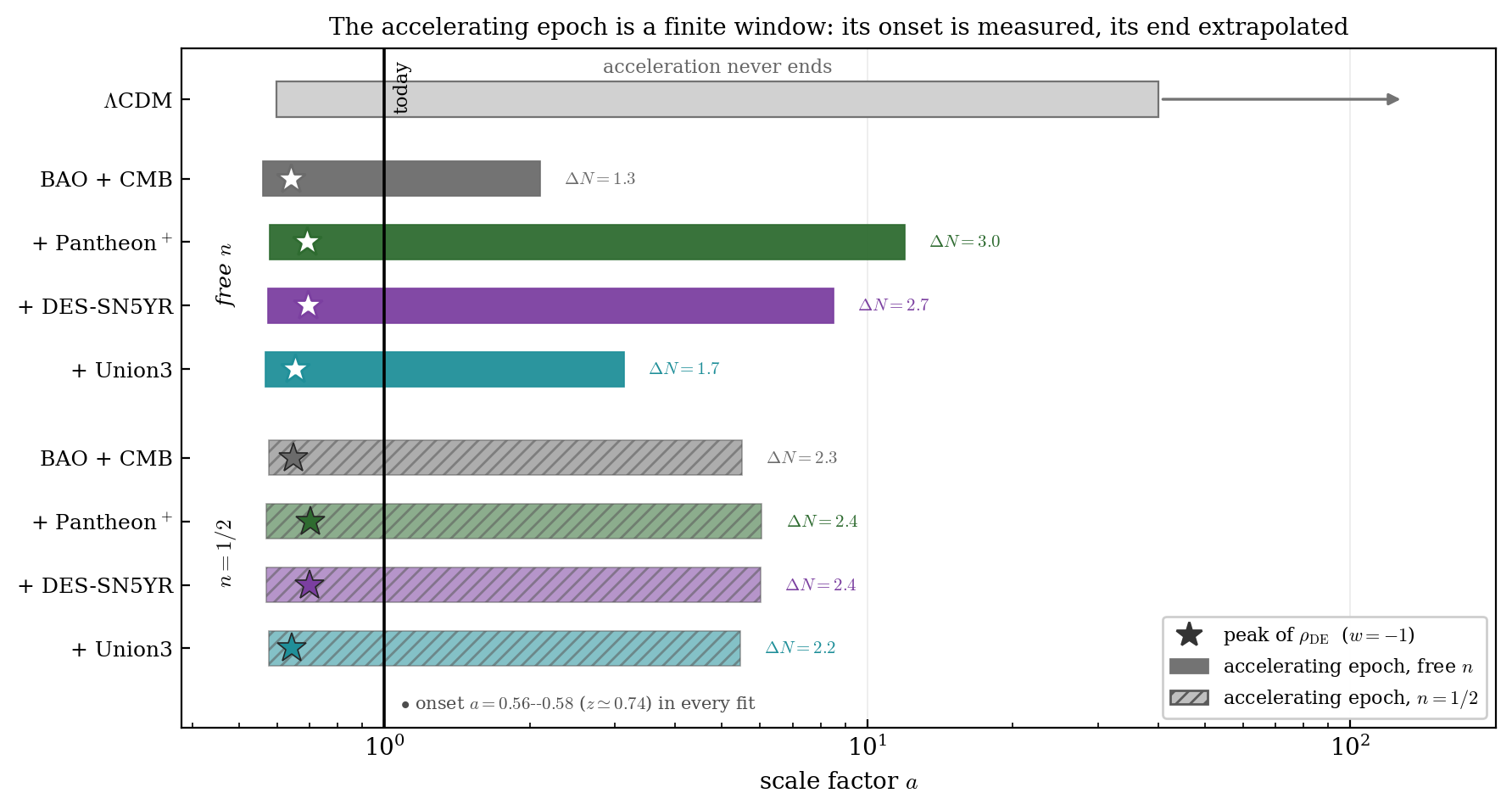}
\caption{\small The accelerating epoch implied by each fit, on a logarithmic scale
factor axis with the present epoch marked. Stars indicate the maximum of
$\rho_{\rm DE}$, which coincides with the phantom crossing. The onset of
acceleration is remarkably stable across every data combination and both
models, $a\simeq0.56$--$0.58$ ($z\simeq0.74$), because it is fixed by the
low-redshift distances that the data actually constrain. The end of
acceleration is not: it varies by a factor of almost six among the free-rate fits, 
since it requires extrapolating the decay of negative pressure over several
future e-folds. Fixing $n=1/2$ removes most of that spread, which reflects the
imposed common rate rather than a measured precision. In every case the
accelerating epoch is a finite window of $\Delta N\simeq1.3$--$3.0$ e-folds
containing the present epoch, in contrast to $\Lambda$CDM (top), for which it
never ends.}
\label{fig:eventtimeline}
\end{figure*}

The fitted histories can be expressed in terms of physical events rather
than only equation-of-state coefficients. All reported relaxation best fits
have $-1<w_0<0$, placing phantom crossing in the past.  
For the linear family the present displacement and the crossing displacement
are $X_0=3w_0/n$ and $X_c=-3/n$, so their ratio is
\begin{equation}
{X_0}/{X_c}=|w_0|
 \label{eq:w-position}
\end{equation}
independently of $n$. The present equation of state is therefore \emph{the
fractional position of the universe between compositional equilibrium and the
phantom crossing}, measured in the logarithmic composition coordinate. That
$|w_0|<1$ is the statement that the crossing lies in the past; that
$|w_0|\simeq0.85$ is the statement that we have covered only about $15\%$ of
the composition interval that separated us from equilibrium at the moment of
crossing. 
At $a_c$ given by
Eq.~\eqref{eq:acLinear}, the DE density reaches its maximum:
it grew before crossing and has decreased since. Its ratio to matter still
increases, because its subsequent dilution remains slower than that of
pressureless matter. The end of the phantom phase therefore need not end
the accelerated epoch.

Acceleration instead depends on the total pressure. Neglecting radiation
at the late epochs considered here, the deceleration parameter is
\begin{equation}
 q(a)=-1-\frac{H'}{H}
 =\frac12\left[1+3w(a)\frac{r(a)}{1+r(a)}\right].
 \label{eq:observational-q}
\end{equation}
The end of the currently accelerating phase is the future root $a_{\rm end}>1$
of $q=0$, equivalently
\begin{equation}
 r(a_{\rm end})\left(3|w_0|a_{\rm end}^{-n}-1\right)=1.
 \label{eq:acceleration-end}
\end{equation}
The abundance matters as well as the equation of state: setting $w=-1/3$
alone would be exact only in the limit of complete DE domination. If that
limit is already a good approximation, it gives the useful estimate
$a_{\rm end}\simeq(3|w_0|)^{1/n}$. The exact root in
Eq.~\eqref{eq:acceleration-end} accounts for the remaining matter contribution.

Elapsed cosmic times follow from
\begin{equation}
\begin{aligned}
 t_0-t_c&=H_0^{-1}\int_{a_c}^{1}\frac{\dd a}{aE(a)},\\
 t_{\rm end}-t_0&=H_0^{-1}\int_{1}^{a_{\rm end}}\frac{\dd a}{aE(a)}.
\end{aligned}
 \label{eq:observational-event-times}
\end{equation}
using, in the same late-time approximation,
\begin{equation}
 E^2(a)=\Omega_m a^{-3}+(1-\Omega_m)a^{-3}
 \exp\!\left[\frac{3w_0}{n}(a^{-n}-1)\right].
 \label{eq:late-time-E}
\end{equation}
Table~III evaluates these relations using the unrounded 
best-fit parameters. Radiation is negligible for these
integrals; uncertainty in the fitted parameters and in the assumed future
continuation is the important limitation.

\begin{table*}[t]
\centering
\renewcommand{\arraystretch}{1.20}
\begin{tabular}{llccccc}
\toprule
data & model & $z_c$ & $t_0-t_c$\,[Gyr] & $a_{\rm beg}$ & $a_{\rm end}$ &
$t_{\rm end}-t_0$\,[Gyr]\\
\midrule
BAO $+$ CMB      & free $n$ & $0.562$ & $5.6$ & $0.560$ & $2.10$  & $15$\\
$+$ Pantheon$^+$ & free $n$ & $0.446$ & $4.7$ & $0.578$ & $11.94$ & $92$\\
$+$ DES-SN5YR    & free $n$ & $0.442$ & $4.7$ & $0.575$ & $8.50$  & $71$\\
$+$ Union3       & free $n$ & $0.527$ & $5.3$ & $0.566$ & $3.14$  & $27$\\
\midrule
BAO $+$ CMB      & $n=1/2$  & $0.543$ & $5.4$ & $0.576$ & $5.51$  & $50$\\
$+$ Pantheon$^+$ & $n=1/2$  & $0.421$ & $4.5$ & $0.570$ & $6.05$  & $53$\\
$+$ DES-SN5YR    & $n=1/2$  & $0.427$ & $4.6$ & $0.570$ & $6.02$  & $53$\\
$+$ Union3       & $n=1/2$  & $0.552$ & $5.5$ & $0.576$ & $5.47$  & $50$\\
\bottomrule
\end{tabular}
\caption{Physical epochs obtained from the relaxation best fits of
Table~\ref{tab:modelComparison}. The
scale factor is unity today; $z_c$ is the phantom crossing and the maximum of
$\rDE$, quoted as a lookback time, while $a_{\rm beg}$ and $a_{\rm end}$
bracket the accelerating epoch and $t_{\rm end}-t_0$ is measured forward from
the present. These are extrapolations of the fitted histories, not confidence
limits, and the spread across data combinations is not an uncertainty
interval. Note the contrast between the onset of acceleration, which is stable
to a few per cent across every row, and its end, which varies by a factor of
almost six.} 
\label{tab:epochs}
\end{table*}
The supernova-inclusive free-rate best fits put the density maximum at
$z_c\simeq0.44$--$0.53$, roughly $4.7$--$5.3$ billion years ago. Their 
predicted end of acceleration varies much more, from approximately $27$ to
$92$ billion years in the future. Fixing the rate to $n=1/2$ brings the 
supernova-inclusive future predictions to about $50$--$53$ billion years,
when the scale factor is approximately $5.5$--$6$ times its present value.
The closer agreement in the restricted model reflects its imposed common
rate; it should not be interpreted as a measured precision on the lifetime
of acceleration. See Fig.~\ref{fig:eventtimeline}.

This contrast has a physical origin. The crossing lies within the recent
history sampled by the distance measurements, while the end of acceleration
requires extrapolating the decay of negative pressure over several future
e-folds. Even the approximate relation
$\ln a_{\rm end}\simeq\ln(3|w_0|)/n$ shows why changes in the rate can
produce large changes in the future epoch. Confidence intervals for that
epoch require propagating the joint parameter constraints, rather than
varying $n$ and $w_0$ independently or using only their best-fit values.

The predicted transition is an end to acceleration, not an end to expansion.
The DE component can retain most of the energy density while losing the
negative pressure that drives acceleration. It ultimately dilutes like
matter at fixed relative abundance, and the spatially flat background
approaches $a\propto t^{2/3}$. The finite accelerated era therefore gives
way to an indefinitely expanding, decelerating future, with no big rip
and no future cosmological event horizon under this continuation.

\paragraph*{Reformulating the coincidence problem.} Why the present-day amplitudes of energy
densities in pressureless matter and dark energy are comparable, despite having a  different 
cosmological evolution? The coincidence problem is particularly severe for $\Lambda$CDM,
in which there is no finite equilibrium: when 
expressed in terms of the quantity $r$ \eqref{eq:rOmega} at the root of our composition variable, the
question is why $r \simeq {\cal O}(1)$ within the 
infinite interval spanning all the positive
real axis. Our framework might ameliorate the issue. With a stable equilibrium  the trajectory is a finite transition between $r\ll1$ and
$r=\rinf$, and the present epoch sits at a determinate position along it ---
quantified, by Eq.~\eqref{eq:w-position}, by the measured value of $w_0$.
The question changes from ``why is $r\sim1$ on the way to $r\to\infty$?'' to
``why are we early in a finite relaxation?'', which is a  
 milder question to address.

What we can say quantitatively is 
 that the present epoch is not located at an
arbitrary point of the history but close to an event the dynamics itself
singles out. The maximum of $\rDE$ --- equivalently the phantom crossing, a
level set of the flow --- occurred $\Delta N_{\rm peak}\simeq0.35$--$0.45$ 
e-folds ago in every fit (Table~\ref{tab:epochs}). Acceleration began at 
$a\simeq0.56$--$0.58$, some $0.55$--$0.58$ e-folds ago, and will end after a
total window of $\Delta N_{\rm acc}\simeq1.3$--$3.0$ e-folds; the present
epoch therefore lies between $18\%$ and $44\%$ of the way through it. The 
coincidence, in this language, is that we observe the universe within
half an e-fold of the maximum of the dark-energy density. That is an
order-unity accident rather than a selection from an unbounded interval, and
in this sense  the runaway part of the problem is
removed.

\section{Dynamics of cosmological fluctuations}
\label{sec_cosflu}

So far we have studied dark-sector relaxation at the level of the
homogeneous evolution, treating $X$ as a coarse-grained background
composition variable. The physical interpretation in terms of relaxation becomes more useful -- and much more interesting -- 
if it also guides the description of cosmological fluctuations. We do so in this Section.

An equation of state
specified only as a function of time determines the background pressure,
but does not specify how that pressure responds to a local disturbance.
Perturbations can be described by supplementing such a parametrization
with additional assumptions, as in effective prescriptions such as
the parametrized post-Friedmann approach~\cite{FangHuLewis2008}.
But our motivation for studying compositional relaxation is to give our assumptions
a  solid physical interpretation. So we connect the evolution of the homogeneous state to
the response of its composition disturbances, invoking tools from linear response theory.

\smallskip
Our starting point is the neighbourhood of compositional equilibrium,
where a smooth attractive flow takes the form
$ X'=-\Gamma(X)$,
$ \Gamma(X)=nX+\mathcal O(X^2)$, $ n>0
 $. 
Onsager's regression hypothesis relates the mean regression of small
spontaneous fluctuations to the macroscopic relaxation of small prepared
departures from equilibrium~\cite{Onsager1R,Onsager2R}. We use this idea as
a guiding principle: long-wavelength composition disturbances should
inherit the near-equilibrium background relaxation rate. 
(Notice that we do
 not
assume a thermal origin for cosmological perturbations or establish
thermodynamic equilibrium of the dark sector.)
In what comes next, we  distinguish a  regression requirement from any additional constitutive assumptions
used to characterize fluctuations at small scales.

\smallskip

In Section \ref{sec:pert-eqs}
 we first identify a
regular, gauge-invariant composition perturbation and construct a minimal response near
equilibrium following standard
regression hypothesis. This is a  consistent  regime  where the background and long-wavelength
disturbances share a relaxation rate. Their relative disturbances have a regular
 acoustic response, and common density perturbations can cluster
gravitationally. In Section  \ref{sec:pert-pointwise} we  examine 
a simple hypothesis governing 
the continuation of the same constitutive
law along all of the cosmological trajectory -- not only nearby equilibrium. Its limitations provide 
requirements for extending the description to the accelerating and
phantom epochs -- the subject of Section \ref{sec:pert-stability}.

\subsection{Composition disturbances and a minimal regression requirement}
\label{sec:pert-eqs}

For the effective fluid realization considered here, we assume Einstein
gravity, separately conserved pressureless matter and DE, and no
anisotropic stress. We work in conformal Newtonian gauge (see e.g. \cite{MalikWandsR}
for a review),
\begin{equation}
 \mathrm ds^2=a^2(\eta)
 \left[-(1+2\Psi)\mathrm d\eta^2
 +(1-2\Phi)\delta_{ij}\mathrm dx^i\mathrm dx^j\right],
 \label{eq:pert-metric}
\end{equation}
and define $\mathcal H=aH$ and
\begin{equation}
 \delta_i=\frac{\delta\rho_i}{\bar\rho_i},\qquad
 \delta X=\delta_{\rm DE}-\delta_m.
 \label{eq:pert-deltaX}
\end{equation}
Bars denote background densities. Under $N\to N+T$, the density
perturbations transform as $\delta_m\to\delta_m+3T$ and
$\delta X\to\delta X+3wT$. The combination
\begin{equation}
 \mathcal X\equiv\delta X-w\delta_m
 =\delta_{\rm DE}-(1+w)\delta_m
 \label{eq:pert-Xcal}
\end{equation}
is therefore gauge invariant, and can be regarded as a natural
definition of perturbation of the coarse-grained quantity $X$. $\mathcal X$ measures the composition displacement
on uniform-matter-density slices. In particular,
$\mathcal X\to\delta_{\rm DE}$ at $w=-1$: using matter as the density
reference avoids dividing by the DE dilution rate, which vanishes at
phantom crossing. Hence, at the level of definition, the
quantity $\mathcal X$ is well behaved at phantom crossing: it is the gauge-invariant variable we alluded to in the first part of Section~\ref{sec:relaxation-model}.

The momentum fluctuation variable is built along the same criteria. Let $\theta_i$ be
the velocity divergence of fluctuations, and  define
\begin{equation}
 U_{\rm DE}\equiv(1+w)\theta_{\rm DE},\qquad
 \Pi\equiv\frac{\delta p_{\rm DE}}{\bar\rho_{\rm DE}}
             -w\delta_{\rm DE}.
 \label{eq:pert-momentum-pressure}
\end{equation}
Here $U_{\rm DE}$ is proportional to the divergence of momentum density,
rather than a velocity obtained by dividing that momentum by the DE
enthalpy. Conservation gives
\begin{align}
 \dot\delta_{\rm DE}
 &=-U_{\rm DE}+3(1+w)\dot\Phi-3\mathcal H\Pi,
 \label{eq:pert-continuity}
 \\
 \dot U_{\rm DE}
 &=-\mathcal H(1-3w)U_{\rm DE}
   +k^2\frac{\delta p_{\rm DE}}{\bar\rho_{\rm DE}}
   +(1+w)k^2\Psi,
 \label{eq:pert-euler}
 \\
 \dot\delta_m&=-\theta_m+3\dot\Phi,
 \qquad
 \dot\theta_m=-\mathcal H\theta_m+k^2\Psi.
 \label{eq:pert-matter}
\end{align}
Dots denote conformal-time derivatives. These equations contain no
division by $1+w$, hence they are regular at phantom crossing. 
 Notice that for regular  modes whose velocity
potentials remain finite in the long-wavelength limit, the momentum
terms are of order $k^2/\mathcal H^2$.
The metric potentials, in addition, are determined by the
linearized Einstein equations sourced by the two components.

Subtracting the previous 
continuity equations we obtain
\begin{equation}
 \delta X'=-3\Pi+3w\Phi'
 -\frac{U_{\rm DE}-\theta_m}{\mathcal H}.
 \label{eq:pert-composition-general}
\end{equation}
Together with Eq.~\eqref{eq:pert-Xcal}, this implies the exact identity
\begin{equation}
 \mathcal X'=-3\Pi-w'\delta_m
 -\frac{U_{\rm DE}-(1+w)\theta_m}{\mathcal H}.
 \label{eq:pert-regression-general}
\end{equation}
The metric derivatives cancel.  
Near equilibrium, the background flow requires
$w=\Gamma/3=nX/3+\mathcal O(X^2)$ and
$w'=-\Gamma_Xw$.

At this stage,  we impose a \emph{regression hypothesis}: linearizing neighbouring homogeneous trajectories
suggests the  requirement
\begin{equation}
 \mathcal X'=-n\mathcal X
 +\mathcal O(X\mathcal X)
 +\mathcal O(k^2/\mathcal H^2),
 \label{eq:pert-regression-target}
\end{equation}
where the last term denotes gradient contributions to the linear
equation. 
In other words, near the equilibrium ($X \ll 1$)the evolution of long
wavelength fluctuations ${\cal X}$ should be  the same as the one of the background composition variable $X$. 

This hypothesis gives a well defined prescription for obtaining a consistent set of near-equilibrium dynamical fluctuations. 
 At the limiting scaling state $X=0$, one has $w=w'=0$ and
$\mathcal X=\delta X$. A simple manipulation of Eq.~\eqref{eq:pert-regression-general}
then shows that the regression requirement \eqref{eq:pert-regression-target} fixes the leading
long-wavelength pressure response to be
\begin{equation}
 \frac{\delta p_{\rm DE}}{\bar\rho_{\rm DE}}
 =\frac n3(\delta_{\rm DE}-\delta_m).
 \label{eq:asymptotic_relative_pressure}
\end{equation}
As a further assumption, we make a \emph{constitutive hypothesis}.
We assume  that the algebraic-type response \eqref{eq:asymptotic_relative_pressure}
also describes finite-wavelength disturbances near equilibrium,
without additional derivative terms, or extra contributions. In this case,
pressure gradients  oppose relative composition disturbances.
We can distinguish adiabatic and non-adiabatic components.

At the adiabatic level, a common perturbation satisfying
\begin{equation}
 \delta_{\rm DE}=\delta_m,\qquad
 \theta_{\rm DE}=\theta_m
 \label{eq:asymptotic_common_mode}
\end{equation}
has no composition displacement and no pressure perturbation.
At the scaling state the conservation equations preserve these
equalities. The corresponding mode evolves as a single dust component and
-- in
the spatially flat matter--DE universe -- it has the usual sub-horizon
growing solution $\delta_{\rm DE}=\delta_m\propto a$ with constant
gravitational potential. Compositional equilibrium therefore permits
gravitational clustering even when the component labelled DE dominates
the energy budget.

The non-adiabatic  displacement mode can be examined independently of the metric.
In the scaling regime, define
\begin{equation}
 D\equiv\delta_{\rm DE}-\delta_m,\qquad
 V\equiv\frac{\theta_{\rm DE}-\theta_m}{\mathcal H}
 \label{eq:pert-relative-variables}
\end{equation}
Since $\mathcal H'/\mathcal H=-1/2$ for matter-like expansion,
Eqs.~\eqref{eq:pert-continuity}--\eqref{eq:pert-matter} give
\begin{equation}
 D'=-nD-V,\qquad
 V'=-\frac12V+\frac{n k^2}{3 \mathcal H^2}D.
 \label{eq:pert-relative-first-order}
\end{equation}
Eliminating $V$ yields
\begin{equation}
 D''+\left(n+\frac12\right)D'
 +\left(\frac n2+\frac{n k^2}{3 \mathcal H^2}\right)D=0.
 \label{eq:pert-relative-equilibrium-wave}
\end{equation}
For $n>0$, both the damping and restoring coefficients are positive.
For regular long-wavelength modes, $V=\mathcal O(q^2)$ and
$D'=-nD+\mathcal O(q^2)$, recovering again the background regression law at large scales. But at short
wavelengths the relative mode is acoustic, with
\begin{equation}
 c_{\rm comp}^{\,2}=\frac n3>0.
 \label{eq:pert-equilibrium-speed}
\end{equation}
The scale $c_{\rm comp}k\sim aH$ marks competition between acoustic
evolution and expansion. See also the general analysis in \cite{Hu:1998kj}.

This 
 response is non-adiabatic. In fact,  for
 the limiting scaling solution we have
$c_a^2=\dot p_{\rm DE}/\dot\rho_{\rm DE}=0$, whereas
(recall Eq.~\eqref{eq:asymptotic_relative_pressure})
\begin{equation}
 \delta p_{\rm nad}=\frac n3\bar\rho_{\rm DE}
 (\delta_{\rm DE}-\delta_m).
 \label{eq:asymptotic_relative_entropy}
\end{equation}
The acoustic coefficient is therefore distinct from the adiabatic
background derivative. It controls the response to displacement from
the equilibrium composition.

Equations~\eqref{eq:pert-regression-target} and
\eqref{eq:pert-relative-equilibrium-wave} give a
near-equilibrium fluctuation content of the proposal. Interestingly,  background relaxation,
regression of composition disturbances, a consistent acoustic response,
and common gravitational clustering coexist within one minimal fluid
prescription, controlled by a single relaxation parameter, $n$. 
Questions about the microscopic origin and the kinetic stability of
an underlying theory go outside the present description, that only  exploits standard hypothesis about systems near
the equilibrium. 

Notice also
that the long-wavelength limit of Eq.~\eqref{eq:pert-relative-equilibrium-wave} has a simple structure,
which suggests something
interesting occurring at our benchmark value $n=1/2$. Its
characteristic polynomial at $k\to0$ factorises exactly,
\begin{equation}
  \lambda^{2}+\Big(n+\tfrac12\Big)\lambda+\frac{n}{2}
  \;=\;\big(\lambda+n\big)\Big(\lambda+\tfrac12\Big),
  \label{eq:factorisation}
\end{equation}
with discriminant $\big(n-\tfrac12\big)^{2}\ge0$. The relative composition
mode is therefore never oscillatory at long wavelength, whatever the relaxation
rate. It decays through two channels, $D\propto e^{-nN}$ and
$D\propto e^{-N/2}$. The first is the regression mode, relaxing at the
background rate; the second is the decay of the relative velocity, since
$\mathcal{H}\propto a^{-1/2}$ in the matter-like expansion of the scaling
state. For adiabatic initial conditions the velocity branch is suppressed,
$V=\mathcal{O}(q^{2})$, and the regression law of Eq.~\eqref{eq:pert-regression-target} is
recovered; an initial relative velocity excites the second branch instead.

The two rates coincide at $n=1/2$, where the mode is critically damped and
\begin{equation}
  D(N)\;=\;\big(c_{1}+c_{2}N\big)\,e^{-N/2}.
  \label{eq:criticaldamping}
\end{equation}
This value also separates two asymptotic behaviours. For $n<1/2$ the regression
mode is the slower of the two and dominates at late times, so the background
relaxation law is what survives; for $n>1/2$ the relative-velocity mode decays
more slowly and eventually dominates, even when initially subdominant. The
supernova-inclusive fits of Sec.~\ref{sec:observables} straddle this boundary,
with $n=0.38$ and $0.43$ on the regression-dominated side and $n=0.65$ on the
other. The square-root benchmark is thus  distinguished at the
level of fluctuations and not only as a fit of background quantities. It is the critically damped point
of the composition mode, at which compositional relaxation is resonant with the
velocity decay of the asymptotic background. We stress that this does not
derive the value $n=1/2$ from an underlying theory. It identifies what would be
special about it, within the near-equilibrium description that the present
framework controls.

\subsection{Continuing the constitutive law along the trajectory: a minimal proposal}
\label{sec:pert-pointwise}

What happens when we leave the equilibrium regime?
The near-equilibrium construction suggests a simple generalization of the relations discussed in Section \ref{sec:pert-eqs}. 
Keeping things as simple as possible,
we
promote the background composition relation to a pointwise constitutive relation:
\begin{equation}
 p_{\rm DE}= w(X)\rho_{\rm DE},
 \label{eq:pert-constitutive}
\end{equation}
This is a new  hypothesis about the pressure response away
from equilibrium -- but it arises naturally as a generalization of the discussion in Section \ref{sub_compo}. An ordinary barotropic fluid has $p=p(\rho)$. Instead, through Eq.~\eqref{eq:pert-constitutive} 
 the DE pressure also depends on the matter environment and
vanishes at a particular density ratio. Perturbing 
Eq.~\eqref{eq:pert-constitutive} gives (recall that  $ w=\Gamma/3$)
\begin{equation}
 \delta p_{\rm DE}
 =w\delta\rho_{\rm DE}
 +\frac{\Gamma_X}{3}\bar\rho_{\rm DE}
 (\delta_{\rm DE}-\delta_m).
 \label{eq:pert-deltap}
\end{equation}
Equivalently,
\begin{equation}
 \frac{\delta p_{\rm DE}}{\bar\rho_{\rm DE}}
 =w\delta_{\rm DE}+\frac{\Gamma_X}{3}\delta X,
 \label{eq:pert-closure-over-rho}
\end{equation}
so that, using definition \eqref{eq:pert-momentum-pressure},
$ \Pi={\Gamma_X}\delta X/3
$.

Under these assumptions,
the density equations discussed in Section~\ref{sec:pert-eqs} become
\begin{align}
 \delta_{\rm DE}'&=-\frac{U_{\rm DE}}{\mathcal H}
 +3(1+w)\Phi'-\Gamma_X\delta X,
 \nonumber\\
 \delta_m'&=-\frac{\theta_m}{\mathcal H}+3\Phi',
 \label{eq:pert-twocont}
\end{align}
and hence
\begin{equation}
 \delta X'=-\Gamma_X\delta X+3w\Phi'
 -\frac{U_{\rm DE}-\theta_m}{\mathcal H}.
 \label{eq:pert-master}
\end{equation}
At leading long-wavelength order this reduces to
\begin{equation}
 \delta X'=-\Gamma_X\delta X+3w\Phi',
 \label{eq:pert-superhorizon-raw}
\end{equation}
where the metric term reflects the slicing dependence of $\delta X$.
Using $w'=-\Gamma_Xw$, 
Eq.~\eqref{eq:pert-regression-general} gives
\begin{equation}
 \mathcal X'=-\Gamma_X\mathcal X
 -\frac{U_{\rm DE}-(1+w)\theta_m}{\mathcal H}.
 \label{eq:pert-composition-pointwise}
\end{equation}
For regular long-wavelength disturbances,
\begin{equation}
 \mathcal X'=-\Gamma_X\mathcal X
 +\mathcal O(k^2/\mathcal H^2).
 \label{eq:pert-superhorizon}
\end{equation}
Hence, even far away from the equilibrium 
the pointwise constitutive relation \eqref{eq:pert-superhorizon} correctly reproduces the linearized
separation law of neighbouring homogeneous trajectories,
see Eq.~\eqref{eq:pert-regression-target}. For
$\Gamma=nX$, the long-wavelength regression rate remains $n$
along the entire history. 

\paragraph*{Regularity at phantom crossing.}
\label{sec:pert-crossing}
\label{sec:pert-standard}
Interestingly,
the composition variable remains a useful evolution coordinate at
phantom 
crossing: while $\rho_{\rm DE}'=0$ at $w=-1$, one has $X'=-3w=3$. In fact,
matter continues to dilute and therefore provides a regular density
reference even at extrema of dark energy evolution. In the conventional adiabatic/non-adiabatic
decomposition~\cite{MalikWandsR},
\begin{equation}
 \delta p_{\rm DE}=c_a^2\delta\rho_{\rm DE}+\delta p_{\rm nad},
 \qquad
 c_a^2=w-\frac{w'}{3(1+w)},
 \label{eq:pert-split}
\end{equation}
the background law gives
\begin{equation}
 c_a^2=w+\frac{\Gamma_Xw}{3(1+w)},
 \label{eq:pert-ca2}
\end{equation}
and Eq.~\eqref{eq:pert-deltap} implies
\begin{equation}
 \delta p_{\rm nad}=\frac{\Gamma_X}{3}\bar\rho_{\rm DE}S,
 \qquad
 S\equiv\frac{\delta_{\rm DE}}{1+w}-\delta_m
       =\frac{\mathcal X}{1+w}.
 \label{eq:pert-nad}
\end{equation}
For a generic crossing and perturbation, both parts
of Eq.~\eqref{eq:pert-nad} can diverge,
but their singular terms cancel exactly in the total pressure
perturbation of Eq.~\eqref{eq:pert-split}. Eq.~\eqref{eq:pert-deltap} remains finite when
the density perturbations and $\Gamma_X$ are finite. The momentum
equations also remain regular. This separates the degeneration of
the conventional variables from physical instability, in the spirit
of regular effective-fluid and post-Friedmann
descriptions~\cite{KunzSapone2006,FangHuLewis2008}.

\smallskip
We stress that this regularity property is a welcome feature
of our compositional framework, but
it does not establish a healthy microscopic crossing for 
the fields that constitute dark energy. The compositional variables
do not remove the microscopic crossing obstructions of particular
theories, see e.g.~\cite{Vikman2005}. Scenarios 
are known to exist, though,
with a safe phantom crossing, as discussed in the Introduction.

\subsection{The limits of a global pointwise closure: a gradient instability}
\label{sec:pert-stability}

The simple constitutive relation \eqref{eq:pert-constitutive} for arbitrary deviations from equilibrium -- although simple and well motivated by its reliable
continuation along the path towards $X\to0$ -- is not free of problems. 
At short wavelengths, the leading density-gradient matrix for
the perturbation vector
 $(\delta_m,\delta_{\rm DE})$ results
\begin{equation}
 \mathbf C=
 \begin{pmatrix}
 0&0\\
 -\Gamma_X/3&w+\Gamma_X/3
 \end{pmatrix}.
 \label{eq:pert-gradient-matrix}
\end{equation}
Its nonzero eigenvalue is
\begin{equation}
 c_{\rm comp}^{\,2}=w+\frac{\Gamma_X}{3}
 =\frac{\Gamma+\Gamma_X}{3}.
 \label{eq:pert-waveeq}
\end{equation}
The associated principal equation contains
$\delta_{\rm DE}''+(k^2/\mathcal H^2)c_{\rm comp}^{\,2}
\delta_{\rm DE}$, together with density mixing and terms of lower
gradient order. Negative $c_{\rm comp}^{\,2}$ therefore signals a
short-wavelength gradient instability -- even though all coefficients
can remain finite at phantom crossing as we have learned above.

On the branch $\Gamma<0$, strict gradient stability requires
\begin{equation}
 \Gamma_X>|\Gamma|.
 \label{eq:pert-stabcond}
\end{equation}
Equivalent forms are
\begin{equation}
 \frac{\mathrm d\ln|\Gamma|}{\mathrm dX}<-1,
 \qquad w'>3w^2,
 \qquad R>|X|\quad(X<0),
 \label{eq:pert-stabforms}
\end{equation}
where $R=\Gamma_X X/\Gamma$ is the local power-law index of the
force. For the exact linear law,
\begin{equation}
 \begin{aligned}
 c_{\rm comp}^{\,2}&=\frac n3(1+X),\\
 c_{\rm comp}^{\,2}>0
 &\ \Longleftrightarrow\ -1<X<0\\
 &\ \Longleftrightarrow\ n>3|w(a)|\quad(X<0).
 \end{aligned}
 \label{eq:pert-stablinear}
\end{equation}
Thus the near-equilibrium negative-pressure branch has a consistent, restoring
gradient response only in proximity of compositional equilibrium. This property is lost
 at $X=-1$, and then for all $|X|>1$.
For $0<n<3$, this occurs before reaching the phantom value
$X_c=-3/n$ when moving away from equilibrium towards the past.
The limitation is therefore not confined to the NEC-violating phase. We checked that the instability is present for the fits of Section \ref{sec:observables}.
For comparison, a constant-$w$ barotropic fluid has
$c_s^2=w$: the positive compositional contribution $n/3$
improves the consistency of  its acoustic response, but is not able to compensate arbitrarily
large negative pressure along the continued linear trajectory (see also \cite{Valiviita2008}).

\paragraph*{A general restriction on continuation into the past.} Going beyond our linear model,
a nonlinear $\Gamma(X)$ can change the size of the stable domain,
but it cannot provide a regular, gradient-stable continuation of
the same pointwise law along a negative-pressure branch to
arbitrarily early scale factors. Allowing the marginal condition
$\Gamma_X+\Gamma\geq0$, define $y=-1/\Gamma>0$. Along the
background trajectory,
\begin{equation}
 y'=\left(-\frac1\Gamma\right)'
 =-\frac{\Gamma_X}{\Gamma}
 =\frac{\Gamma_X}{|\Gamma|}\geq1.
 \label{eq:nogoderiv}
\end{equation}
Integrating backwards from a stable reference epoch $N_b$ gives
\begin{equation}
 \frac1{|\Gamma(N)|}
 \leq\frac1{|\Gamma(N_b)|}-(N_b-N).
 \label{eq:nogobound}
\end{equation}
The right-hand side reaches zero after
\begin{equation}
 \Delta N_{\rm stab}\leq\frac1{|\Gamma(N_b)|}
 =\frac1{3|w(N_b)|}.
 \label{eq:nogoDN}
\end{equation}
A finite negative-pressure branch therefore cannot remain regular
and gradient stable beyond this backward interval. Its continuation
must lose stability, become singular in its constitutive variables,
or leave the assumptions of the pointwise description. Moreover,
if the branch extends to $X\to-\infty$, the same inequality gives
\begin{equation}
 |\Gamma(X)|\geq|\Gamma(X_b)|e^{X_b-X},\qquad
 \int_{-\infty}^{X_b}\frac{\mathrm dX}{|\Gamma(X)|}
 \leq\frac1{|\Gamma(X_b)|}.
 \label{eq:pert-past-integral-bound}
\end{equation}
The accumulated backward interval in $N$ is finite. Thus a
regular, gradient-stable pointwise negative-pressure branch can not
 cover the full past evolution to $a\to0$.
Nevertheless, this restriction concerns the global constitutive continuation \eqref{eq:pert-constitutive} -- it does not concern
 the near-equilibrium construction  of Section \ref{sec:pert-eqs}, which remains intact. It also shows
that modifying only the background function $\Gamma$ is not a
general cure: additional pressure-response physics is required
outside the domain where the minimal local relation is adequate.

\subsection{Pressure response beyond the near-equilibrium domain: a possible remedy}
\label{sec:pert-response-extension}

The out-of-equilibrium results and instability problems discussed above suggest useful requirements for an
extension of the program that goes beyond the minimal
pointwise constitutive relation of \eqref{eq:pert-constitutive}. We seek
a setup which preserves the fitted background trajectory,
remains regular at phantom crossing, provides a consistent spatial
response, and recovers the original Onsager-type composition regression and
restoring pressure near $X=0$. These conditions guide our search, yet 
without uniquely determining a microscopic model.

For the linear family, we consider the  modification
\begin{equation}
 \begin{aligned}
 \delta p_{\rm DE}
 &=\delta p_{\rm DE}^{(0)}
   +A\bar\rho_{\rm DE}\mathcal X,\\
 \frac{\delta p_{\rm DE}^{(0)}}{\bar\rho_{\rm DE}}
 &=w\delta_{\rm DE}+\frac n3(\delta_{\rm DE}-\delta_m) 
 \end{aligned}
 \label{eq:repair_general}
\end{equation}
for a quantity $A$ to be determined.
The added term is gauge invariant. It modifies the response to
a composition disturbance, while leaving the selected unperturbed
history unchanged. It also vanishes for the adiabatic mode
$\mathcal X=0$. 
Expanding Eq.~\eqref{eq:repair_general},
\begin{equation}
 \frac{\delta p_{\rm DE}}{\bar\rho_{\rm DE}}
 =\left(w+\frac n3+A\right)\delta_{\rm DE}
 -\left[\frac n3+A(1+w)\right]\delta_m.
 \label{eq:repair_expanded}
\end{equation}
Its principal acoustic eigenvalue is $w+n/3+A$. Within such specific Ansatz, the choice $A=-w$ is uniquely selected when
requiring such eigenvalue to equal the composition equilibrium value $n/3$
at every epoch (recall Eq.~\eqref{eq:pert-equilibrium-speed}). It gives
\begin{equation}
 \frac{\delta p_{\rm DE}}{\bar\rho_{\rm DE}}
 =\frac n3\delta_{\rm DE}
 +\left[w(1+w)-\frac n3\right]\delta_m,
 \label{eq:repair_closure}
\end{equation}
and
\begin{equation}
 c_{\rm comp}^{\,2}=\frac n3>0.
 \label{eq:repair_speed}
\end{equation}
For $0<n\leq3$, this principal acoustic speed is at most luminal
within the fluid prescription. Notice that no additional background
fit parameter is introduced in this modified prescription -- but a new response assumption has
been made, yet compatible with the linear relaxation hypothesis. The pressure law is retained along the selected history
and is no longer imposed pointwise on all configurations.

Since the modified pressure remains finite, the momentum equations
\eqref{eq:pert-continuity}--\eqref{eq:pert-matter} remain regular at
$w=-1$. The short-wavelength gradient instability of the original
closure is removed at the level of its principal acoustic
coefficient. The correction also vanishes as $w\to0$, recovering
the near-equilibrium pressure law of Eq. ~\eqref{eq:asymptotic_relative_pressure}:
\begin{equation}
 \delta p_{\rm DE}\longrightarrow
 \frac n3\bar\rho_{\rm DE}(\delta_{\rm DE}-\delta_m).
 \label{eq:repair_asymptotic}
\end{equation}
\paragraph*{Regression and neighbouring homogeneous patches.}
However, the modification of constitutive relation \eqref{eq:repair_general} also changes the long-wavelength dynamics.
Equation~\eqref{eq:pert-regression-general} now yields
\begin{equation}
 \mathcal X'=-(n+3A)\mathcal X
 +\mathcal O(k^2/\mathcal H^2).
 \label{eq:repair_regression}
\end{equation}
For $A=-w$ and $w=nX/3$, this becomes
\begin{equation}
 \begin{aligned}
 \mathcal X'&=-n(1-X)\mathcal X
 +\mathcal O(k^2/\mathcal H^2)\\
 &=-n\mathcal X+\mathcal O(X\mathcal X)
 +\mathcal O(k^2/\mathcal H^2),
 \end{aligned}
 \label{eq:repair_near_equilibrium}
\end{equation}
to be confronted with Eq.~\eqref{eq:pert-superhorizon}.
Thus the prescribed background and the long-wavelength
disturbances recover the same regression rate near equilibrium,
but they do not share it away from equilibrium.

This distinction is important for the separate-universe
interpretation. Preserving one homogeneous history does not
preserve the family of neighbouring homogeneous histories.
For $A=-w$, the correction survives at $k/(aH)\to0$
and changes the
entropy response. Although the adiabatic mode is preserved,
composition disturbances can no longer be identified with
neighbouring solutions of the original one-variable flow without
additional homogeneous dynamics.

\paragraph*{A response with distinct long- and short-wavelength limits?}
If the original separate-universe law is retained, a possible
direction is a response that vanishes as $k/(aH)\to0$ while
approaching $A=-w$ at short wavelengths. An illustrative form is
\begin{equation}
 \begin{gathered}
 A(a,k)=-w(a)F(q/q_c),\qquad q=\frac{k}{aH},\\
 F(y)=\mathcal O(y^2)\quad(y\to0),\qquad
 F(y)\to1\quad(y\to\infty),
 \end{gathered}
 \label{eq:repair_scale_dependent}
\end{equation}
where $q_c$ is a dimensionless transition scale. A regular, bounded function
$F$ recovers the original leading long-wavelength regression and
the positive short-wavelength acoustic coefficient. Because $A$
vanishes as $w\to0$, it also preserves the near-equilibrium response. It remains to be demonstrated, in a concrete model, if we can establish acceptable dynamics
at intermediate wavelengths. 

\subsection{From a near-equilibrium response to a physical completion}
\label{sec:pert-outlook}

In this Section \ref{sec_cosflu} we  established
 a minimal
near-equilibrium response connecting the macroscopic relaxation
rate to composition regression and acoustic restoration, while
allowing common gravitational clustering. The obstruction to a
global pointwise continuation then identifies which part of that
description must be extended to cover the observed accelerating
epoch sufficiently far from equilibrium. The pressure modifications illustrate how preserving the
near-equilibrium limit can guide this extension, and how
long-wavelength consistency constrains apparently successful
short-wavelength repairs.

A route complementary to ours  is to embed the prescribed background in
a theory whose equations already determine the perturbative
response. VCDM, for example, has auxiliary fields and constraints
that allow an evolving effective DE background without an
independent propagating DE scalar mode~\cite{DeFeliceDollMukohyama2020,MukohyamaNoui2019,VCDMH0R,VCDMWeakenR,AroraR,LsCDMR,LsCDMConstraintsR}.
Its matter and metric fluctuations remain dynamical, and the
constrained gravitational equations supply their response.
In such a realization, the composition flow organizes the
background, while the chosen gravitational theory determines
whether and how the constitutive interpretation is reproduced.

\smallskip

The next theoretical task is therefore to identify a realization
that explains the relaxation clock and pressure response,
establishes kinetic and gradient stability within its regime of
validity, and consistently connects neighbouring homogeneous
patches to finite-wavelength fluctuations. Its predictions for
growth and lensing can then distinguish realizations with similar
expansion histories. The near-equilibrium composition response
provides definite conditions for this programme;
the full cosmological trajectory provides the setting in which
its extensions can be tested.

\section{Conclusions and Outlook}
\label{sec:discussion}

So, what is the future of our universe? In this
work we addressed our initial question by exploring
a framework where acceleration is  part of a
relaxation process towards equilibrium, corresponding to stationary relative abundance of matter
and DE. The  composition variable we analyzed
removes the common dilution of separately conserved components
and connects their unequal dilution rates directly to the DE
equation of state. The composition--force plane organizes phantom
crossings, density extrema and scaling equilibria within a common
geometric description. 

Near a smooth, nondegenerate attractive fixed point
corresponding to the endpoint of evolution, linear
relaxation supplies the leading response. Our minimal global
hypothesis extends this law to the whole background trajectory,
$X'=-nX$, giving $w=w_0a^{-n}$. The parameter $n$ specifies
the relaxation rate, while $w_0$ fixes the present state along the
trajectory; in cosmic time the instantaneous rate is $nH$.
The global continuation is an additional modelling assumption,
whose value lies in connecting the observed epoch to definite
past and future behaviour with the same two parameters.

Importantly, 
the near-equilibrium description also provides a constructive
starting point for fluctuations. Guided by Onsager's regression
hypothesis, we require long-wavelength composition disturbances
to relax with the background rate. Conservation then fixes their
leading pressure response at the scaling state. A minimal local
continuation gives relative disturbances a positive acoustic
coefficient, $c_{\rm comp}^{\,2}=n/3$, while common matter--DE
perturbations remain free to cluster gravitationally. The same
parameter therefore connects background relaxation, regression
and acoustic restoration within the chosen constitutive relation. Regression is used as a guiding principle. These fluid
results can give a physical completion definite conditions to preserve,
like in the Kadanoff-Martin approach \cite{KadanoffMartin}.

The background likelihood analysis supports further investigation
of this picture. Within the adopted likelihoods, nonzero relaxation
improves the fit over constant $w$ at a nominal
$2.6$--$3.6\sigma$, and the general family fits comparably to CPL.
The benchmark $n=1/2$ is compatible with all three supernova
combinations at the nominal $68\%$ profile-likelihood level and
retains comparable fit quality with one fewer DE parameter.
The modest information-criterion gains and overlap with its
motivating data support this benchmark as an economical model
for further tests, without establishing a fundamental rate.

A direct observational target is the relaxation law itself.
Constant fractional relaxation predicts a constant
$n_{\rm eff}=-\mathrm d\ln|w|/\mathrm d\ln a$. Our null test
finds no significant variation, although present sensitivity is
limited. Improved supernova and BAO measurements can sharpen this
test and search for nonlinear response without fixing a particular
value of $n$. They can thereby assess the evolution law over the
observed interval, while its scaling future remains a conditional
extrapolation.

Extending the fluctuation response beyond the neighbourhood of
equilibrium is an interesting theoretical task. The gauge-invariant
composition disturbance  we defined remains regular at phantom crossing,
where the DE density ceases to provide a useful reference clock. Choosing the benchmark $n=1/2$ has interesting
consequences for the evolution of fluctuations. 
Regularity does not ensure stability, however. We established that,
with separate conservation, the pointwise compositional pressure
law cannot support a regular, gradient-stable negative-pressure
branch throughout the past. This restriction identifies the limits
of that constitutive continuation and guides its extension.
The pressure response requires physical freedom beyond the
instantaneous density ratio to cover the full trajectory.


Theoretical and observational developments of our results can proceed
together. Background reconstructions can constrain the relaxation
rate and test its constancy; physical realizations can explain
the relaxation clock and extend the near-equilibrium pressure
response to the observed epoch. Growth and lensing can then test
those realizations. Compositional relaxation connects these tasks
through a common physical question: whether today's acceleration
is a stage in an approach towards shared scaling, and how the
relaxation of that composition governs both the background
evolution and its fluctuations.

\acknowledgments
\noindent
It is a pleasure to thank  Ivonne Zavala for input and discussions.
Standard AI models have been used for brainstorming, refining parts of the presentation, and assisting with the  numerical procedure described in Appendix A. GT is partially supported by the STFC grant [UKRI4243], by the Royal Society grant IES R3 43186, and by the
Leverhulme Trust grant RF-2026-166 9.

\appendix
\setcounter{section}{0}

\section{Technical details of the background analysis}
\label{app:method}

This appendix explains the numerical implementation of the fits presented in Sec.~\ref{sec:observables}. The
data sets, their cuts and the supernova marginalization are described there;
what follows is the numerical construction.

\paragraph{Background evolution}

Each history enters through one function, the dark-energy density in units of
its present value,
\begin{equation}
\begin{aligned}
 f_{\rm DE}(a)&=\exp\!\left[-3\int_1^a\bigl(1+w\bigr)\frac{\dd a'}{a'}\right]\\
 &=\begin{cases}
 1, & \LCDM,\\[2pt]
 a^{-3(1+w_0)}, & w{\rm CDM},\\[2pt]
 a^{-3}\exp\!\left[\frac{3w_0}{n}(a^{-n}-1)\right], & \text{relaxation},\\[2pt]
 a^{-3(1+w_0+w_a)}e^{-3w_a(1-a)}, & \text{CPL}.
 \end{cases}
\end{aligned}
 \label{eq:a-fde}
\end{equation}
and the expansion rate follows from flatness,
\begin{equation}
\begin{aligned}
 E^2(a)&=\Omega_ra^{-4}+\Om a^{-3}\\
 &\quad +(1-\Om-\Omega_r)f_{\rm DE}(a),\\
 \Omega_r&=\frac{\Om}{1+z_{\rm eq}}.
\end{aligned}
 \label{eq:a-E}
\end{equation}
with $z_{\rm eq}=2.5\times10^4\,\Om h^2(T_{\rm CMB}/2.7\,{\rm K})^{-4}$ and
$T_{\rm CMB}=2.7255\,$K. Here $\Om$ is the total pressureless component, cold
dark matter plus baryons. The first two lines of \eqref{eq:a-fde} are the
$n\to0$ limit of the third, so $\LCDM$ and $w$CDM are obtained as constrained
fits within the relaxation family rather than as separate models. The
relaxation case is evaluated as $-3\ln a+3w_0\,\mathrm{expm1}(-n\ln a)/n$ for
$|n|>10^{-7}$ and as $-3(1+w_0)\ln a$ below, so that the limit is continuous.

\paragraph{Distances and sound horizons}

Comoving and Hubble distances are $D_M(z)=(c/H_0)\int_0^z\dd z'/E$ and
$D_H=c/H$, with $D_V=(zD_HD_M^2)^{1/3}$; the integral uses $80$-node
Gauss--Legendre quadrature. The BAO observables are these divided by the drag
horizon, taken from the standard fitting formulas \cite{BOSS:2014hhw}
\begin{equation}
\begin{gathered}
 r_d=\frac{55.154\exp\!\left[-72.3(\omega_\nu+0.0006)^2\right]}
      {\omega_m^{0.25351}\omega_b^{0.12807}}\ {\rm Mpc},\\
 \omega_\nu=0.00064.
\end{gathered}
 \label{eq:a-rd}
\end{equation}
corresponding to $\sum m_\nu=0.06\,$eV held fixed. Since \eqref{eq:a-rd}
involves only pre-recombination quantities, it is independent of the
DE model: BAO provides a ruler whose length is set by early-time
physics and whose apparent size is governed entirely by the late-time
expansion.

The distance priors are
\begin{equation}
 R=\frac{\sqrt{\Om}H_0D_M(z_*)}{c},
 \qquad
 \ell_A=\frac{\pi D_M(z_*)}{r_s(z_*)},
 \label{eq:a-priors}
\end{equation}
with mean $(1.7502,301.471,0.02236)$, standard deviations
$(0.0046,0.0895,0.00015)$ and correlations $\rho_{R\ell_A}=0.46$,
$\rho_{R\omega_b}=-0.66$, $\rho_{\ell_A\omega_b}=-0.33$. The decoupling
redshift uses the Hu--Sugiyama formula \cite{Hu:1995en}
\begin{equation}
\begin{gathered}
 z_*=1048\bigl(1+0.00124\,\omega_b^{-0.738}\bigr)(1+g_1\omega_m^{g_2}),\\
 g_1=\frac{0.0783\,\omega_b^{-0.238}}{1+39.5\,\omega_b^{0.763}},\qquad
 g_2=\frac{0.560}{1+21.1\,\omega_b^{1.81}},
\end{gathered}
\end{equation}
and the sound horizon is integrated directly,
\begin{equation}
\begin{aligned}
 r_s(z_*)&=\frac{c}{H_0}\int_0^{a_*}
 \frac{\dd a}{a^2E\sqrt{3(1+R_b)}},\\
 R_b&=\frac{3\omega_b}{4\omega_\gamma}a,\\
 \omega_\gamma&=\frac{3}{4}\,\frac{(T_{\rm CMB}/2.7\,{\rm K})^4}{31500},
\end{aligned}
\end{equation}
with $180$ quadrature nodes. Note that $r_d$ is obtained from a fitting
formula while $r_s(z_*)$ is integrated; the two enter different likelihoods
and are never combined.

\paragraph{Likelihood evaluation and minimization}

Each covariance is Cholesky-factorized once, ${\bf C}=\bm{\mathsf L}\bm{\mathsf
L}^{\mathsf T}$, and the analysis works with whitened residuals
$\bm{\mathsf L}^{-1}(\bm m-\bm d)$, whose sum of squares is $\chi^2$. For the
supernovae the offset is first projected out and the result whitened with the
symmetric square root of the supplied inverse covariance, so that the sum of
squares reproduces the marginalized $\chi^2$ of Eq.~(4.1) exactly. The vector
passed to the minimizer is the concatenation of $13$ BAO, $3$ distance-prior
and $N_{\rm SN}$ supernova entries. Writing the likelihood this way allows a
least-squares algorithm to exploit the Jacobian structure, which matters
because the profiles below require of order $10^3$ minimizations.

Minimization uses a bounded trust-region least-squares method with
Jacobian-based scaling and tolerances $10^{-11}$, over
$\Om\in[0.15,0.45]$, $H_0\in[50,90]\,\kms$, $\omega_b\in[0.018,0.026]$, with
$w_0\in[-1.5,-0.3]$ and $n\in[\epsilon,3]$ for the relaxation family and
$w_0\in[-2,0]$, $w_a\in[-4,2]$ for CPL. All best fits lie in the interior.
The CPL likelihood has more than one local minimum on BAO $+$ CMB, so those
fits are repeated from a $4\times4$ grid of starting points and the lowest
retained; the other models converge from a single start.

One numerical point deserves mention. The $n\to0$ limit must be approached
with a small positive $\epsilon$ (we use $10^{-8}$) and the out-of-range test
must be a strict inequality: with $\epsilon=0$ and a non-strict test, every
fit holding $n$ at exactly zero --- that is, $\LCDM$, $w$CDM and the endpoint
of the profile in $n$ --- is rejected at the boundary and the minimizer
returns its starting point. We have verified that $\chi^2$ is continuous
across the limit. Relatedly, the bound $n\ge\epsilon$ imposes rather than
tests the attractiveness of the composition fixed point; a measurement of the
sign of $n$ requires relaxing it to negative values.

\paragraph{Profile likelihoods}

The parameter of interest is fixed to each value of a grid ($73$ points on
$n\in[\epsilon,2.4]$ in Fig.~3) and all remaining parameters are re-minimized,
giving $\chi^2_{\min}(n)$. Each fit starts from the converged solution of the
previous grid point, so that the minimizer follows the degeneracy ridge
continuously; this matters on BAO $+$ CMB, where the $(w_0,n)$ valley is long
and curved. Interval endpoints come from linear interpolation of the crossings
with $\Delta\chi^2=1$ and $3.84$; two-dimensional contours use $2.30$ and
$6.18$.

A profile likelihood is not a marginal posterior. The two agree when the
likelihood is close to Gaussian in the parameters and can differ when the
degeneracy is strongly curved, which is the case for the geometry-only
$(w_0,n)$ constraint. The supernova-inclusive intervals, where that degeneracy
is broken, are correspondingly the more robust.

\paragraph{Derived quantities}

The crossing follows in closed form, $a_c=|w_0|^{1/n}$. The accelerating epoch
is bracketed by the roots of $1+3w_{\rm tot}=0$ with
$w_{\rm tot}=wr/(1+r)$ and $r(a)=[(1-\Om-\Omega_r)/\Om]\,a^3f_{\rm DE}(a)$,
located on a logarithmic grid of $2\times10^5$ points over
$a\in[10^{-4},10^3]$ and refined by bisection. Elapsed times use
$t_2-t_1=H_0^{-1}\int_{a_1}^{a_2}\dd a/(aE)$ by adaptive quadrature, with
$H_0^{-1}=977.792/(H_0/\kms)\,$Gyr. Present-day cosmographic quantities are
obtained by finite differencing $\ln E$, with step $\delta z=10^{-4}$ for
$q_0$ and a second difference of step $2\times10^{-3}$ for $j_0$.

\paragraph{Validation}

Four checks were performed. (i) Fitting flat $\LCDM$ to each supernova
compilation alone returns $\Om=0.332\pm0.018$ for Pantheon$^+$ and
$\Om=0.356^{+0.028}_{-0.026}$ for Union3, against published values
$0.334\pm0.018$ and $0.356^{+0.028}_{-0.026}$; DES gives $0.330\pm0.015$,
differing from the $0.352\pm0.017$ of the 2024 release because the public data
now ship the Dovekie recalibration. (ii) The CPL fit on BAO $+$ CMB reproduces
the published DESI DR2 $w_0w_a$ constraint. (iii) That same fit,
$(w_0,w_a,H_0,\Om)=(-0.437,-1.670,63.90,0.351)$, agrees with a full
Boltzmann analysis of Planck 2018 $+$ DESI DR2 in VCDM, which gives
$(-0.45,-1.61,63.9,0.349)$; this is the most direct evidence that the
compression retains the information relevant to late-time dark energy.
(iv) For flat $\LCDM$ the finite-difference cosmography returns the jerk parameter \cite{Borghetto2026}
$j_0=1.000174$ against the exact value $1$, and $q_0$ to better than
$2\times10^{-4}$. Quadrature convergence was checked by doubling all node
counts, with $\chi^2$ stable at the $10^{-6}$ level.

\section{A direct null test of constant relaxation}
\label{app_null}

\begin{table*}[t]
\centering
\small
\setlength{\tabcolsep}{4pt}
\begin{tabular}{lcccccc}
\toprule
data & $n_<$ & $n_>$ & $\Delta n$ & $\Delta\chi^2_{\Delta n=0}$ & $\beta$ & $\Delta\chi^2_{\beta=0}$ \\
\midrule
+ Pantheon$^+$ & \shortstack{$-0.06$\\$[-0.45,0.31]$} & \shortstack{$0.90$\\$[0.47,1.44]$} & \shortstack{$\phantom{-}0.97$\\$[0.22,1.81]$} & 1.69 & \shortstack{$\phantom{-}1.20$\\$[-0.05,2.67]$} & 0.92\\
\addlinespace[4pt]
+ DES-SN5YR    & \shortstack{$\phantom{-}0.29$\\$[-0.10,0.68]$} & \shortstack{$0.58$\\$[0.20,1.03]$} & \shortstack{$\phantom{-}0.29$\\$[-0.42,1.06]$} & 0.16 & \shortstack{$\phantom{-}0.30$\\$[-0.83,1.63]$} & 0.06\\
\addlinespace[4pt]
+ Union3       & \shortstack{$\phantom{-}0.66$\\$[0.12,1.20]$}  & \shortstack{$0.65$\\$[0.23,1.16]$} & \shortstack{$\phantom{-}0.00$\\$[-0.89,0.94]$} & 0.00 & \shortstack{$-0.35$\\$[-1.67,1.21]$} & 0.06\\
\bottomrule
\end{tabular}
\caption{Null test of constant relaxation. Best-fit values with 68\% profile-likelihood
intervals for the two-bin rates of Eq.~\eqref{eq:neff_bins}, their difference
$\Delta n$, and the drift parameter $\beta$. The columns $\Delta\chi^2$ give the
cost of imposing the null hypothesis, which is the linear relaxation model in
both cases. All data combinations include DESI DR2 BAO and the Planck distance
priors.}
\label{tab:nulltest}
\end{table*}

\begin{figure*}[t]
\centering
\paperfigure{1.4in}{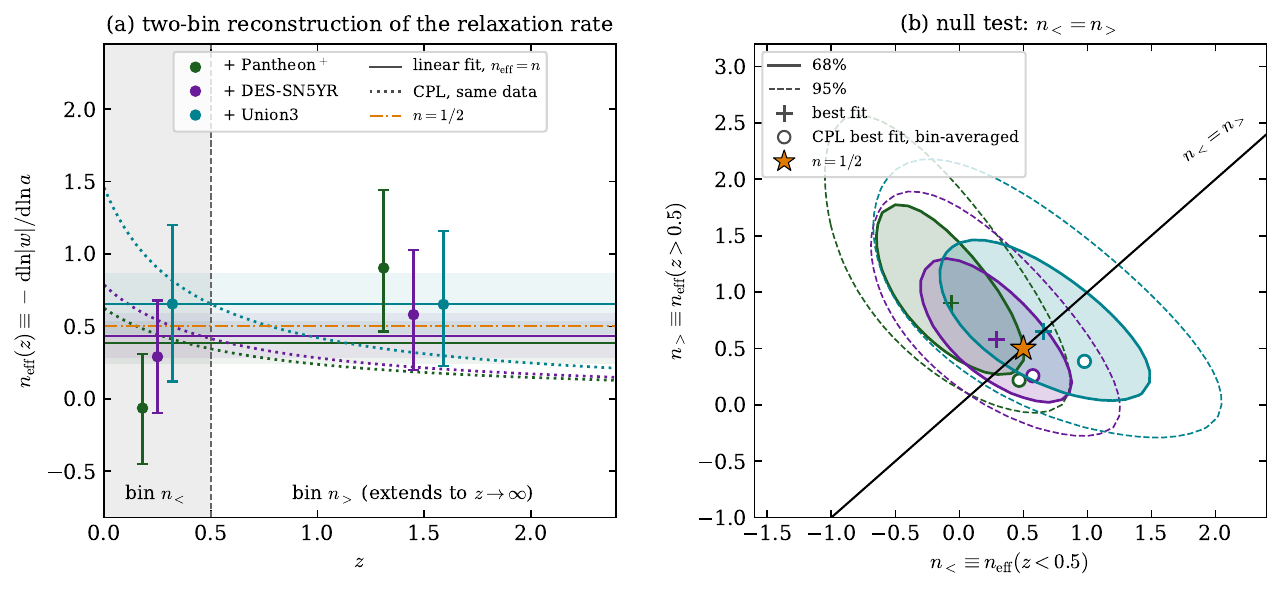}
\caption{\emph{Left:} two-bin reconstruction of the relaxation rate
$n_{\rm eff}=-{\rm d}\ln|w|/{\rm d}\ln a$, with 68\% profile intervals (points,
offset horizontally for clarity). The lower bin covers $z<0.5$, after the phantom
crossing; the upper bin covers all $z\ge0.5$. Solid lines and bands show the
constant rate of the linear fits, and dotted curves show $n_{\rm eff}(z)$ for CPL
fitted to the same data. CPL forces $n_{\rm eff}\to0$ in the past, because $w$
approaches the constant $w_0+w_a$. \emph{Right:} 68\% and 95\% two-dimensional
profile contours in $(n_<,n_>)$. Linear relaxation is the diagonal; the star marks
$n=1/2$. Open circles show the CPL best fits averaged over the same bins (upper
bin evaluated over $0.5<z<2.33$).}
\label{fig:nulltest}
\end{figure*}

The analysis of Section \ref{sec:observables}  tests the relaxation model through its overall goodness of
fit. It does not isolate the property that defines it. By Eq.~\eqref{eq:crossingDirection}, the
logarithmic slope of the equation of state is the local slope of the
composition force,
\begin{equation}
  n_{\rm eff}(z) \equiv -\frac{{\rm d}\ln|w|}{{\rm d}\ln a} = \Gamma_X\big(X(z)\big),
\end{equation}
and the linear law $\Gamma=nX$ is  the statement that this quantity is
the same at every epoch. Constancy of $n_{\rm eff}$ makes no reference to the
value of the rate or to the present state $w_0$, so it can be tested without
assuming either. In the coordinate $\ln|w|$ versus $\ln a$ the null hypothesis
is a straight line.

To test the null-channel hypothesis
we therefore replace the single rate by a piecewise-constant one,
\begin{equation}
\begin{aligned}
 w(a)&=w_0\exp\!\left[\int_a^1 n_{\rm eff}(a')\,\frac{{\rm d}a'}{a'}\right],\\
 n_{\rm eff}(z)&=\begin{cases}
 n_< , & z<z_s,\\
 n_> , & z\ge z_s.
 \end{cases}
\end{aligned}
 \label{eq:neff_bins}
\end{equation}
with $z_s=0.5$. This choice places the bin edge close to the fitted phantom
crossing, $z_c\simeq0.45$--$0.53$ (Table~\ref{tab:epochs}), so the test compares the
relaxation rate after and before the maximum of $\rho_{\rm DE}$. The upper bin
extends to all higher redshifts. The early-time behaviour therefore remains
that of a linear flow, with exponential suppression of dark energy. The sign of $w$ is
preserved by construction. Linear relaxation is recovered exactly for $n_<=n_>$,
and the null statistic is
\begin{equation}
  \Delta n \equiv n_> - n_< .
\end{equation}
Two bins are the resolution of present data: $n_{\rm eff}$ is a derivative of
the reconstructed history, and with three bins the uncertainty on each rate
already exceeds unity. As a one-parameter complement we also consider a linear
drift of the rate over the observed range,
$n_{\rm eff}(z)=n+\beta\,\min[\ln(1+z),\ln(1+z_{\max})]$ with $z_{\max}=2.33$,
the highest DESI redshift. The null hypothesis is $\beta=0$. On BAO+CMB alone, $w_0$ and the two rates are
essentially unconstrained in this extended model, and we quote results only for
the supernova-inclusive combinations.

The results are collected in Table~\ref{tab:nulltest} and
Fig.~\ref{fig:nulltest}. All three data combinations pass the null test. The
largest departure occurs for Pantheon$^+$, which prefers a faster relaxation
before the crossing than after it, $\Delta n = 0.97^{+0.84}_{-0.75}$. Imposing
$\Delta n=0$ costs only $\Delta\chi^2=1.69$, a nominal $1.3\sigma$. For
DES-SN5YR and Union3 the two rates agree to $0.4\sigma$ and to within numerical
precision, respectively. The drift parameter is consistent with zero in every
case, with $\Delta\chi^2_{\beta=0}\le0.92$. Present background data therefore
show no evidence for curvature of $\ln|w|$ in $\ln a$ over $0<z\lesssim2.3$.

The location of the fixed-rate model in this plane is informative. The point
$n_<=n_>=1/2$ lies inside the joint 68\% region for every combination, with
$\Delta\chi^2=2.27$, $0.34$ and $0.57$ for two parameters. For Pantheon$^+$ it
sits on the boundary of that region. The CPL best fits, averaged over the same
bins, lie below the diagonal ($n_< > n_>$). This is a structural property of CPL
rather than a feature of the data: $n_{\rm eff}^{\rm CPL}=w_a a/w$ decreases
monotonically into the past and vanishes as $a\to0$, whereas constant relaxation
keeps it fixed. At present the CPL points are also allowed, with
$\Delta\chi^2=2.83$, $0.74$ and $0.44$. With $\sigma(\Delta n)\simeq0.8$, the
test excludes only gross departures from constant relaxation, roughly
$|\Delta n|\gtrsim2$ at 95\%. Separating a CPL-like drift of
$\Delta n\simeq-0.3$ to $-0.6$ from $\Delta n=0$ would require an uncertainty
about four times smaller.

The null test is therefore currently passed, but not yet demanding. Its value
lies on the fact that it targets the defining property of the model,
independently of the rate and of the present state, and requires no parameter
beyond those already fitted. Because $n_{\rm eff}$ is controlled mainly by the
density of supernovae below $z\simeq1$ and by the BAO lever arm above it,
forthcoming DESI releases and the next generation of supernova samples will
sharpen it directly.

\end{document}